\documentclass[11pt]{article}

\usepackage[latin1]{inputenc}\usepackage{epsfig}\usepackage{amsfonts}

\title{\huge
  Low-Pass Filters, Fourier Series and \\
  Partial Differential Equations }

\author{
  \Large Jorge L. deLyra \\
  Department of Mathematical Physics \\
  Physics Institute \\
  University of São Paulo }

\date{March 4, 2015}

\newcommand{\FFrac}[2]{{\displaystyle\frac{\displaystyle #1}{\displaystyle #2}}}
\newcommand{\Sum}{\displaystyle\sum}

\newcommand{\at}[2]{\left.\rule{0em}{3ex}\right[_{\,#1}^{\,#2}}

\begin{document}\maketitle

\begin{abstract}
  \noindent
  When Fourier series are used for applications in physics, involving
  partial differential equations, sometimes the process of resolution
  results in divergent series for some quantities. In this paper we argue
  that the use of linear low-pass filters is a valid way to regularize
  such divergent series. In particular, we show that these divergences are
  always the result of oversimplification in the proposition of the
  problems, and do not have any fundamental physical significance. We
  define the first-order linear low-pass filter in precise mathematical
  terms, establish some of its properties, and then use it to construct
  higher-order filters. We also show that the first-order linear low-pass
  filter, understood as a linear integral operator in the space of real
  functions, commutes with the second-derivative operator. This can
  greatly simplify the use of these filters in physics applications, and
  we give a few simple examples to illustrate this fact.
\end{abstract}

\section{Introduction}

One of the many important uses of Fourier series is in the role of tools
for the solution of boundary value problems involving partial differential
equations in Cartesian or cylindrical coordinates~\cite{ChurFour}. From a
historical point of view, one may say that this is, in fact, the original
use of these series. In this type of application the final solution of the
boundary value problem is obtained in the form of a Fourier series, and
the real function that gives its coefficients is not immediately available
in closed form. This is not surprising, since the determination of the
function is the very essence of the boundary value problem. It is often
necessary to take derivatives of the solutions obtained, in order to
calculate physical quantities of relevance within the applications, and in
some cases the term-wise differentiation of the Fourier series obtained
for the solution results in a divergent series.

This is usually brought about by overly simplified initial conditions or
boundary conditions, which are used when formulating the physical problem
in order to simplify the sequence of operations leading to the solution in
the form of a Fourier series. Common cases in which this can happen are
the calculation of the acceleration in problems with the vibrating string,
the calculation of the electric field in problems involving the
electrostatic potential within a box and the calculation of the heat flux
density in problems involving the heat equation in solids. Some simple
examples on these lines are examined briefly in
Appendix~\ref{APPexampfilt}, in order to illustrate the main points of
this paper.

In this paper we show that certain low-pass filters can be used to deal
with such situations in a way that changes the mathematics so as to make
the relevant Fourier series converge, while at the same time changing
nothing of importance in the physics involved. The first-order linear
low-pass filter is defined in precise mathematical terms, and several of
its main properties are established, both on the real line and within a
periodic interval. The first-order filter is then used to define
higher-order filters, the use of which not only results in convergent
Fourier series, but in series that also converge faster and to smoother
functions, allowing one to take a few term-by-term derivatives, as needed
within the applications involved.

Although the concept of a low-pass filter originates from an engineering
practice, it can be defined theoretically in precise mathematical terms,
as an operation on real functions. In fact, all the linear low-pass
filters discussed here can be understood as integral operators acting in
the space of integrable real functions. They can be expressed by integrals
on the real line, involving certain kernel functions. They can be defined
both on the whole real line and within a periodic interval such as
$[-\pi,\pi]$, which allows one to write Fourier expansions for the real
functions, in the form

\begin{displaymath}
  f(x)
  =
  \frac{1}{2}\,
  \alpha_{0}
  +
  \sum_{k=1}^{\infty}
  \left[
    \alpha_{k}\cos(kx)
    +
    \beta_{k}\sin(kx)
  \right],
\end{displaymath}

\noindent
where the coefficients are given by

\noindent
\begin{eqnarray*}
  \alpha_{k}
  & = &
  \frac{1}{\pi}
  \int_{-\pi}^{\pi}dx\,
  \cos(kx)f(x),
  \\
  \beta_{k}
  & = &
  \frac{1}{\pi}
  \int_{-\pi}^{\pi}dx\,
  \sin(kx)f(x).
\end{eqnarray*}

\noindent
Since boundary value problems typically involve partial differential
equations within compact domains, it is of particular importance to
determine the action of the low-pass filters on functions defined on
compact intervals. We will also establish how the filters act directly on
the Fourier representations of these functions.

\section{The Low-Pass Filters on the Real Line}

The low-pass filters are defined as operations on real functions leading
to other related real functions. Let us define the simplest such filter,
namely the {\em first-order linear low-pass filter}. Given a real function
$f(x)$ on the real line of the coordinate $x$, of which we require no more
than that it be integrable, we define from it a filtered function
$f_{\varepsilon}(x)$ by

\begin{equation}\label{deffilt}
  f_{\varepsilon}(x)
  =
  \frac{1}{2\varepsilon}
  \int_{x-\varepsilon}^{x+\varepsilon}dx'\,
  f\!\left(x'\right),
\end{equation}

\noindent
where $\varepsilon$ is a strictly positive real constant, usually meant to
be small by comparison to some physical scale, and which we will refer to
as the {\em range} of the filter. One can also define $f_{0}(x)$ by
continuity, as the $\varepsilon\to 0$ limit of this expression, which is
mostly but not always identical to $f(x)$. The transition from $f(x)$ to
$f_{\varepsilon}(x)$ constitutes an operation within the space of real
functions. A discrete version of this operation is known in numerical and
graphical settings as that of taking {\em running averages}. Another
similar operation is known in quantum field theory as {\em block
  renormalization}. What we do here is to map the value of $f(x)$ at $x$
to its average value over a symmetric interval around $x$. This results in
a new real function $f_{\varepsilon}(x)$ that is smoother than the
original one, since the filter clearly damps out the high-frequency
components of the Fourier spectrum of $f(x)$, as will be shown explicitly
in what follows.

The filter can be understood as a linear integral operator acting in the
space of integrable real functions. It may be written as an integral over
the whole real line involving a kernel
$K_{\varepsilon}\!\left(x-x'\right)$ with compact support,

\begin{displaymath}
  f_{\varepsilon}(x)
  =
  \int_{-\infty}^{\infty}dx'\,
  K_{\varepsilon}\!\left(x-x'\right)
  f\!\left(x'\right),
\end{displaymath}

\noindent
where the kernel is defined as
$K_{\varepsilon}\!\left(x-x'\right)=1/(2\varepsilon)$ for
$|x-x'|<\varepsilon$, and as $K_{\varepsilon}\!\left(x-x'\right)=0$ for
$|x-x'|>\varepsilon$. This kernel is a discontinuous even function of
$\left(x-x'\right)$ that has unit integral. If the functions one is
dealing with are defined in a periodic interval such as $[-\pi,\pi]$, then
the integral above has to be restricted to that interval, and the kernel
can be easily expressed in terms of a convergent Fourier series,

\begin{displaymath}
  K_{\varepsilon}\!\left(x-x'\right)
  =
  \frac{1}{2\pi}
  +
  \frac{1}{\pi}
  \sum_{k=1}^{\infty}
  \left[
    \frac{\sin(k\varepsilon)}{(k\varepsilon)}
  \right]
  \cos\!\left[k\left(x-x'\right)\right],
\end{displaymath}

\noindent
where we assume that $\varepsilon\leq\pi$. The calculation of the
coefficients of this series is completely straightforward. The series can
be shown to be convergent by the Dirichlet test, or alternatively by the
monotonicity criterion discussed in~\cite{FTotCPII}. The quantity within
square brackets is known as the sinc function of the variable
$(k\varepsilon)$. In spite of appearances, it is an analytic function,
assuming the value $1$ at zero.

Although it is possible to define the filter of range $\varepsilon$ inside
a periodic interval even if the overall range is larger that the length of
the interval, that is when $\varepsilon>\pi$ in our case here, there is
little point in doing so. The central idea of the filter is that the range
be small compared to the relevant scales of a given problem, and once a
periodic interval is introduces it immediately establishes such a scale
with its length. Therefore we should have at least $\varepsilon\leq\pi$,
and more often $\varepsilon\ll\pi$. We will therefore adopt as a basic
hypothesis, from now on, the condition that the range be smaller than the
length of the periodic interval, whenever we work with periodic functions
within such an interval.

The filter defined above has several interesting properties, which are the
reasons for its usefulness. Some of the most important and basic ones
follow. In every case it is clear that $f(x)$ must be an integrable
function, otherwise it is not even possible to define the corresponding
filtered function.

\begin{enumerate}

\item If $f(x)$ is a linear function on the real line, then
  $f_{\varepsilon}(x)=f(x)$.

\item If $f(x)=x^{n}$ on the real line, then $f_{\varepsilon}(x)$ is a
  polynomial of order $n$, with the coefficient $1$ for the term $x^{n}$.

  Only lower powers of $x$ with the same parity as $n$ appear in this
  polynomial. All the other coefficients contain strictly positive powers
  of $\varepsilon^{2}$, and thus tend to zero when $\varepsilon\to 0$.
  This means that in the $\varepsilon\to 0$ limit the filter becomes the
  identity, in so far as polynomials are concerned.

\item If $f(x)$ is a continuous function, then $f_{\varepsilon}(x)$ is a
  differentiable function.

\item If $f(x)$ is a discontinuous function, then $f_{\varepsilon}(x)$ is
  a continuous function.

\item If $f(x)$ is an integrable singular object such as Dirac's delta
  ``function'', then $f_{\varepsilon}(x)$ is a discontinuous function. In
  fact, the kernel defined above can itself be obtained by the application
  of the filter to a delta ``function''.

\item In the $\varepsilon\to 0$ limit the filter becomes an
  almost-identity operation, in the sense that it reproduces in the output
  function $f_{\varepsilon}(x)$ the input function $f(x)$ almost
  everywhere.

\item At isolated points where $f(x)$ is discontinuous the $\varepsilon\to
  0$ limit of the function $f_{\varepsilon}(x)$ converges to the average
  of the two lateral limits of $f(x)$ to that point.

\item At isolated points where $f(x)$ is non-differentiable the
  $\varepsilon\to 0$ limit of the derivative of the function
  $f_{\varepsilon}(x)$ converges to the average of the two lateral limits
  of the derivative of $f(x)$ to that point.

\item\label{FiltProp09} The filter does not change the definite integral
  of a function that has compact support on the real line.

\end{enumerate}

\noindent
Up to this point we have assumed that $f(x)$ is defined on the whole real
line. If instead of this it is defined within a periodic interval, then we
have a few more properties.

\begin{enumerate}\setcounter{enumi}{9}

\item If $f(x)$ is periodic, then so is $f_{\varepsilon}(x)$, with the
  same period.

\item The filter does not change the average value of a periodic function.
  This means that it does not change the integral of the function over its
  period, and hence that it does not change the Fourier coefficient
  $\alpha_{0}$ of the function.

\item\label{FiltProp12} For periodic functions the effect of the filter on
  the asymptotic behavior of the Fourier coefficients $\alpha_{k}$ and
  $\beta_{k}$ of the function, for $k>0$, is to add an extra factor of $k$
  to the denominator. This is so because the filtered coefficients may be
  written as

  \noindent
  \begin{eqnarray*}
    \alpha_{\varepsilon,k}
    & = &
    \left[
      \frac{\sin(k\varepsilon)}{(k\varepsilon)}
    \right]
    \alpha_{k},
    \\
    \beta_{\varepsilon,k}
    & = &
    \left[
      \frac{\sin(k\varepsilon)}{(k\varepsilon)}
    \right]
    \beta_{k}.
  \end{eqnarray*}

  \noindent
  Once more we see here the presence of the sinc function of the variable
  $(k\varepsilon)$.

\end{enumerate}

\noindent
All these properties can be demonstrated directly on the real line, and
some such demonstrations can be found in Appendix~\ref{APPpropfilt}. For
our purposes here one of the most important properties is the last one,
since it implies that the action of the filter, when represented in the
Fourier series of the real function, is very simple and has the effect of
rendering the filtered series more rapidly convergent than the original
one, since the filtered coefficients contain an extra factor of $1/k$ and
hence approach zero faster than the original ones as we make $k\to\infty$.

The usefulness of the filter in physics applications, and the very
possibility of using it to regularize divergent Fourier series in such
circumstances, stem from two facts related to the mathematical
representation of nature in physics. First, such a representation is
always an approximate one. All physical measurements, as well as all
theoretical calculations, of quantities which are represented by
continuous variables, can only be performed with a finite amount of
precision, that is, within finite and non-zero errors. In fact, not only
this is true in practice, but with the advent of relativistic quantum
mechanics and quantum field theory, it became a limitation in principle as
well. Second, all physical laws are valid within a certain range of
length, time or energy scales. Given any physical measurements or
theoretical calculations, there is always a length or time scale below
which, or an energy scale above which, the measurements and calculation,
as well as the hypotheses behind them, cease to have any meaning.

If we observe that the application of a filter with range parameter
$\varepsilon$ appreciably changes the function, and therefore the
representation of nature that it implements, only at scales of the order
of $\varepsilon$ or smaller, while at the same time resulting in series
with better convergence characteristics for {\em all} non-zero values of
$\varepsilon$, no matter how small, it becomes clear that it is always
possible to choose $\varepsilon$ small enough so that no appreciable
change in the physics is entailed within the relevant scales. We conclude
therefore that it is always possible to filter the real functions involved
in physics applications, in order to have a representation of the physics
in terms of convergent series, without the introduction of any physically
relevant changes in the description of nature and its laws. In fact, many
times it turns out that the introduction of the low-pass filter actually
{\em improves} the approximate representation of nature used in the
applications, rather than harming it in any way, as shown in the examples
discussed in Appendix~\ref{APPexampfilt}.

\subsection{Higher-Order Filters}

Since the first-order filter defined here is linear, one can construct
higher-order filters by simply applying it multiple times to a given real
function. Consider the first-order filter of range $\varepsilon$ written
in terms of the first-order kernel,

\begin{displaymath}
  f_{\varepsilon}^{(1)}(x)
  =
  \int_{-\infty}^{\infty}dx'\,
  K_{\varepsilon}^{(1)}\!\left(x-x'\right)
  f\!\left(x'\right),
\end{displaymath}

\noindent
where the first-order kernel is given, now in full detail, by the
piece-wise description

\noindent
\begin{equation}\label{defK1full}
  %
  \renewcommand{\arraystretch}{2.0}
  \begin{array}{rclcc}
    K_{\varepsilon}^{(1)}\!\left(x-x'\right)
    & = &
    0
    &
    \mbox{for}
    &
    \varepsilon<\left(x-x'\right),
    \\
    K_{\varepsilon}^{(1)}\!\left(x-x'\right)
    & = &
    \FFrac{1}{4\varepsilon}
    &
    \mbox{for}
    &
    \varepsilon=\left(x-x'\right),
    \\
    K_{\varepsilon}^{(1)}\!\left(x-x'\right)
    & = &
    \FFrac{1}{2\varepsilon}
    &
    \mbox{for}
    &
    -\varepsilon<\left(x-x'\right)<\varepsilon,
    \\
    K_{\varepsilon}^{(1)}\!\left(x-x'\right)
    & = &
    \FFrac{1}{4\varepsilon}
    &
    \mbox{for}
    &
    \left(x-x'\right)=-\varepsilon,
    \\
    K_{\varepsilon}^{(1)}\!\left(x-x'\right)
    & = &
    0
    &
    \mbox{for}
    &
    \left(x-x'\right)<-\varepsilon.
  \end{array}
\end{equation}

\noindent
Note that, although this is not important for its operation, at the points
of discontinuity we define the value of the kernel as the average of the
two lateral limits to that point. These are the values to which its
Fourier series converges at these points. With this the kernel can also be
given by the Fourier representation within $[-\pi,\pi]$, if
$\varepsilon\leq\pi$,

\begin{displaymath}
  K_{\varepsilon}^{(1)}\!\left(x-x'\right)
  =
  \frac{1}{2\pi}
  +
  \frac{1}{\pi}
  \sum_{k=1}^{\infty}
  \left[
    \frac{\sin(k\varepsilon)}{(k\varepsilon)}
  \right]
  \cos\!\left[k\left(x-x'\right)\right].
\end{displaymath}

\noindent
Using the representation in terms of an integral operator it is easy to
compose two instances of the first-order filter in order to obtain a
second-order one, with range $2\varepsilon$,

\noindent
\begin{eqnarray*}
  f_{2\varepsilon}^{(2)}(x)
  & = &
  \int_{-\infty}^{\infty}dx'\,
  K_{\varepsilon}^{(1)}\!\left(x-x'\right)
  f_{\varepsilon}^{(1)}\!\left(x'\right)
  \\
  & = &
  \int_{-\infty}^{\infty}dx'\,
  K_{\varepsilon}^{(1)}\!\left(x-x'\right)
  \int_{-\infty}^{\infty}dx''\,
  K_{\varepsilon}^{(1)}\!\left(x'-x''\right)
  f\!\left(x''\right)
  \\
  & = &
  \int_{-\infty}^{\infty}dx''\,
  \left[
    \int_{-\infty}^{\infty}dx'\,
    K_{\varepsilon}^{(1)}\!\left(x-x'\right)
    K_{\varepsilon}^{(1)}\!\left(x'-x''\right)
  \right]
  f\!\left(x''\right)
  \\
  & = &
  \int_{-\infty}^{\infty}dx''\,
  K_{2\varepsilon}^{(2)}\!\left(x-x''\right)
  f\!\left(x''\right),
\end{eqnarray*}

\noindent
where the second-order kernel with range $2\varepsilon$ is given by the
application of the first-order filter to the first-order kernel,

\begin{displaymath}
  K_{2\varepsilon}^{(2)}\!\left(x-x''\right)
  =
  \int_{-\infty}^{\infty}dx'\,
  K_{\varepsilon}^{(1)}\!\left(x-x'\right)
  K_{\varepsilon}^{(1)}\!\left(x'-x''\right).
\end{displaymath}

\noindent
It is not difficult to show by direct calculation of the integral that
this second-order kernel is given by the piece-wise description

\noindent
\begin{displaymath}
  %
  \renewcommand{\arraystretch}{2.0}
  \begin{array}{rclcc}
    K_{2\varepsilon}^{(2)}\!\left(x-x'\right)
    & = &
    0
    &
    \mbox{for}
    &
    2\varepsilon\leq\left(x-x'\right),
    \\
    K_{2\varepsilon}^{(2)}\!\left(x-x'\right)
    & = &
    \FFrac{1}{2\varepsilon}
    \left(
      1
      -
      \FFrac{x-x'}{2\varepsilon}
    \right)
    &
    \mbox{for}
    &
    0\leq\left(x-x'\right)\leq 2\varepsilon,
    \\
    K_{2\varepsilon}^{(2)}\!\left(x-x'\right)
    & = &
    \FFrac{1}{2\varepsilon}
    \left(
      1
      +
      \FFrac{x-x'}{2\varepsilon}
    \right)
    &
    \mbox{for}
    &
    -2\varepsilon\leq\left(x-x'\right)\leq 0,
    \\
    K_{2\varepsilon}^{(2)}\!\left(x-x'\right)
    & = &
    0
    &
    \mbox{for}
    &
    \left(x-x'\right)\leq-2\varepsilon,
  \end{array}
\end{displaymath}

\noindent
which makes its range explicit. It is also given by the Fourier
representation within $[-\pi,\pi]$, so long as $\varepsilon\leq\pi/2$,

\begin{displaymath}
  K_{2\varepsilon}^{(2)}\!\left(x-x'\right)
  =
  \frac{1}{2\pi}
  +
  \frac{1}{\pi}
  \sum_{k=1}^{\infty}
  \left[
    \frac{\sin(k\varepsilon)}{(k\varepsilon)}
  \right]^{2}
  \cos\!\left[k\left(x-x'\right)\right].
\end{displaymath}

\noindent
The calculation of the coefficients of this series is just as
straightforward as the one for the first-order kernel. Due to the factor
of $k^{2}$ in the denominator, this series is absolutely and uniformly
convergent over the whole periodic interval. The result shown above also
follows from the property of the first-order filter regarding its action
on Fourier expansions, listed as item~\ref{FiltProp12} on
page~\pageref{FiltProp12}, which is demonstrated in
Section~\ref{APPfourcoefs} of Appendix~\ref{APPpropfilt}. Note that,
according to the property listed as item~\ref{FiltProp09} on
page~\pageref{FiltProp09}, which is demonstrated in
Section~\ref{APPinvints} of Appendix~\ref{APPpropfilt}, the first-order
filter does not change the definite integral of the compact-support
function it is applied on, and since
$K_{\varepsilon}^{(1)}\!\left(x-x'\right)$ is an even function with unit
integral and compact support, it follows that
$K_{2\varepsilon}^{(2)}\!\left(x-x'\right)$ is also an even function with
unit integral and compact support, since it is given by the first-order
kernel $K_{\varepsilon}^{(1)}\!\left(x-x'\right)$ filtered by the
first-order filter.

The range of the first-order filter, within which the functions are
significantly changed by it, is given by $\varepsilon$, and if one just
applies the filter twice, as we did above, that range doubles do
$2\varepsilon$. However, one may compensate for this by simply applying
twice the first-order filter with parameter $\varepsilon/2$, thus
resulting in a second-order filter with range $\varepsilon$. In this way
one may define higher-order filters while keeping the relation of the
range to the relevant physical scale constant. For example, we have the
second-order filter with range $\varepsilon$ defined by the kernel

\begin{displaymath}
  K_{\varepsilon}^{(2)}\!\left(x-x''\right)
  =
  \int_{-\infty}^{\infty}dx'\,
  K_{\varepsilon/2}^{(1)}\!\left(x-x'\right)
  K_{\varepsilon/2}^{(1)}\!\left(x'-x''\right).
\end{displaymath}

\noindent
It is immediate to obtain the piece-wise description of this kernel from
that of $K_{2\varepsilon}^{(2)}\!\left(x-x'\right)$,

\noindent
\begin{displaymath}
  %
  \renewcommand{\arraystretch}{2.0}
  \begin{array}{rclcc}
    K_{\varepsilon}^{(2)}\!\left(x-x'\right)
    & = &
    0
    &
    \mbox{for}
    &
    \varepsilon\leq\left(x-x'\right),
    \\
    K_{\varepsilon}^{(2)}\!\left(x-x'\right)
    & = &
    \FFrac{1}{\varepsilon}
    \left(
      1
      -
      \FFrac{x-x'}{\varepsilon}
    \right)
    &
    \mbox{for}
    &
    0\leq\left(x-x'\right)\leq\varepsilon,
    \\
    K_{\varepsilon}^{(2)}\!\left(x-x'\right)
    & = &
    \FFrac{1}{\varepsilon}
    \left(
      1
      +
      \FFrac{x-x'}{\varepsilon}
    \right)
    &
    \mbox{for}
    &
    -\varepsilon\leq\left(x-x'\right)\leq 0,
    \\
    K_{\varepsilon}^{(2)}\!\left(x-x'\right)
    & = &
    0
    &
    \mbox{for}
    &
    \left(x-x'\right)\leq-\varepsilon.
  \end{array}
\end{displaymath}

\noindent
It is equally immediate to obtain the Fourier representation of this
kernel within $[-\pi,\pi]$, which so long as $\varepsilon\leq\pi$ is given
by

\begin{displaymath}
  K_{\varepsilon}^{(2)}\!\left(x-x'\right)
  =
  \frac{1}{2\pi}
  +
  \frac{1}{\pi}
  \sum_{k=1}^{\infty}
  \left[
    \frac{\sin(k\varepsilon/2)}{(k\varepsilon/2)}
  \right]^{2}
  \cos\!\left[k\left(x-x'\right)\right].
\end{displaymath}

\noindent
This procedure can be iterated $N$ times to produce an order-$N$ filter.
One can verify on a case-by-case fashion that such a filter is given by a
piece-wise kernel formed of polynomials of order $N-1$ on $N$ equal-length
intervals between $-N\varepsilon$ and $N\varepsilon$, each interval of
length $2\varepsilon$, with the polynomials connected to each other in a
maximally smooth way. Since the filter of order $N$ is obtained by the
application of the first-order filter to the result of the filter of order
$N-1$, it follows that the kernel of order $N$ is the kernel of order
$N-1$ filtered by the first-order filter. Due to this, and recalling again
the property of the first-order filter regarding its action on Fourier
expansions, listed as item~\ref{FiltProp12} on page~\pageref{FiltProp12},
the Fourier representation of the order-$N$ kernel of range $N\varepsilon$
can be easily written explicitly,

\begin{displaymath}
  K_{N\varepsilon}^{(N)}\!\left(x-x'\right)
  =
  \frac{1}{2\pi}
  +
  \frac{1}{\pi}
  \sum_{k=1}^{\infty}
  \left[
    \frac{\sin(k\varepsilon)}{(k\varepsilon)}
  \right]^{N}
  \cos\!\left[k\left(x-x'\right)\right],
\end{displaymath}

\noindent
so long as $\varepsilon\leq\pi/N$. This expression can be extended to the
case $N=0$, which corresponds to an order-zero filter that has the Dirac
delta ``function'' as its kernel, since the delta ``function'' can be
represented by the divergent series

\noindent
\begin{eqnarray*}
  \delta\!\left(x-x'\right)
  & = &
  K_{0}^{(0)}\!\left(x-x'\right)
  \\
  & = &
  \frac{1}{2\pi}
  +
  \frac{1}{\pi}
  \sum_{k=1}^{\infty}
  \cos\!\left[k\left(x-x'\right)\right],
\end{eqnarray*}

\noindent
as is discussed in detail in~\cite{FTotCPI}. We see in this way that the
first-order kernel $K_{\varepsilon}^{(1)}\!\left(x-x'\right)$ can in fact
be obtained by the application of the first-order filter to the delta
``function'', as is discussed in more detail in Section~\ref{APPactdelta}
of Appendix~\ref{APPpropfilt}.

In this construction the range of the filter increases with $N$, so that
one cannot iterate in this way indefinitely inside the periodic interval
without the range eventually becoming larger than the length of the
interval. However, we may keep the overall range constant at the value
$\varepsilon$ by decreasing the range of the first-order filter at each
level of iteration, that is, by iterating $N$ times the first-order filter
of range $\varepsilon/N$. If we simply exchange $\varepsilon$ for
$\varepsilon/N$ in the expression above we get the order-$N$ kernel with
range $\varepsilon$, written in quite a simple way in terms of its Fourier
expansion,

\begin{displaymath}
  K_{\varepsilon}^{(N)}\!\left(x-x'\right)
  =
  \frac{1}{2\pi}
  +
  \frac{1}{\pi}
  \sum_{k=1}^{\infty}
  \left[
    \frac{\sin(k\varepsilon/N)}{(k\varepsilon/N)}
  \right]^{N}
  \cos\!\left[k\left(x-x'\right)\right],
\end{displaymath}

\begin{figure}[ht]
  \centering
  \fbox{
    \epsfig{file=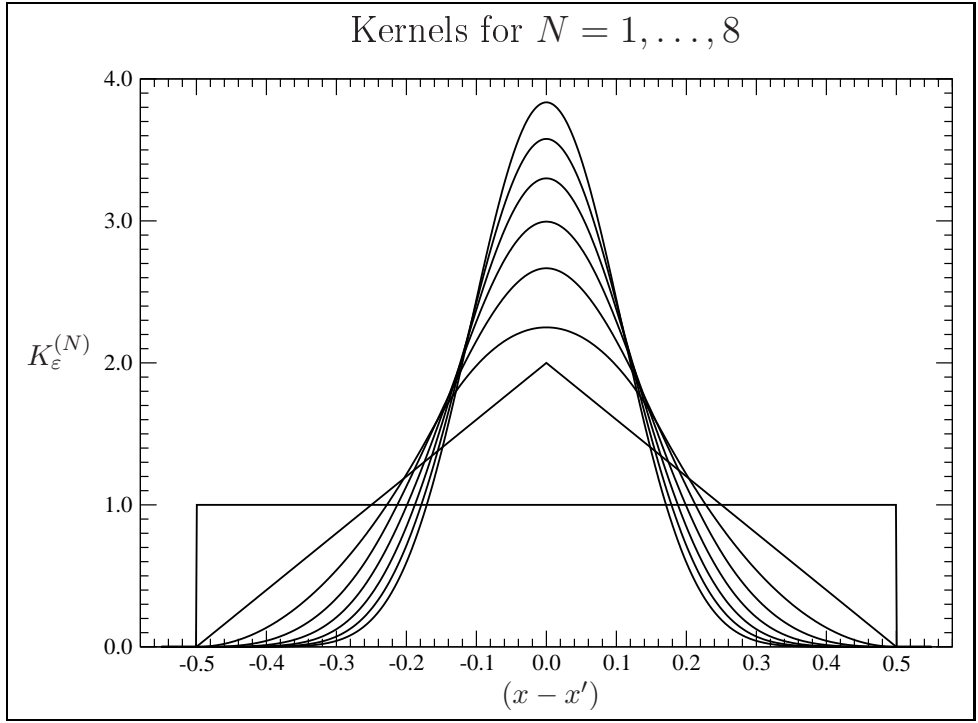,scale=1.0,angle=0}
  }
  \caption{The kernels of the first few lower-order filters with constant
    range $\varepsilon$, obtained via the use of their Fourier series, for
    $N=1,\ldots,8$ and $\varepsilon=0.5$, plotted as functions of
    $\left(x-x'\right)$ over their common support within the periodic
    interval $[-\pi,\pi]$.}
  \label{Fig01}
\end{figure}

\noindent
so long as $\varepsilon\leq\pi$. Since the range is now constant, one can
consider iterations of any order $N$, without any upper bound, even within
the periodic interval. Note that this series converges ever faster as $N$
increases. Note also that it can be differentiated $N-2$ times still
resulting in a absolutely and uniformly convergent series, and $N-1$ times
still resulting in a point-wise convergent series. This is a reflection of
the fact that the polynomials that compose the kernel are connected to
each other in the maximally smooth way. Apart from the case of the
order-zero kernel, which has a divergent Fourier series, the series for
$K_{\varepsilon}^{(1)}\!\left(x-x'\right)$ is the only one which is not
absolutely or uniformly convergent, although it is point-wise convergent.
For $N\geq 2$ all the kernel series are absolutely and uniformly
convergent to functions of differentiability class $C^{N-2}$ everywhere.
The kernels of the filters of the first few orders, with constant range
$\varepsilon$, are shown in Figure~\ref{Fig01}. The program used to plot
this graph is available online~\cite{TarFile}.

As we saw above, the order-$N$ kernels are themselves a good example of
the smoothing action of the filters. As we verified in that case, the use
of higher-order filters will have the effect of introducing more powers of
$k$ in the denominators of the Fourier coefficients $\alpha_{k}$ and
$\beta_{k}$, and hence of making the Fourier series converge faster and to
smoother functions. This will then enable one to take a certain number of
term-wise derivatives of the series, as may be required by the
applications involved. Besides, all this can be done within a small
constant length scale determined by $\varepsilon$, leaving essentially
untouched the description of the physics at the larger scales.

\section{Application in Partial Differential Equations}

Let us now describe how one can use the low-pass filters in boundary value
problems involving partial differential equations. The basic idea is that,
if the solution of a boundary value problem leads to a divergent Fourier
series for some physical quantity, then the correct physical
interpretation of this fact is that the mathematical description of the
physical system being dealt with lacks sufficient realism. This is usually
a problem contained within the initial conditions used, or within the
boundary conditions used, or both. The divergences are always consequences
of singularities contained within these conditions. We therefore use the
filters in order to smooth out the initial or boundary conditions, using
some small range parameter $\varepsilon$ which is suggested by the
relevant physical scales of the physical system. Having done that, we may
then repeat the whole resolution of the boundary value problem. The
solution obtained in this way will then present lesser convergence
problems, and quite probably none at all.

While using this technique, it is useful to keep in mind some basic
mathematical and physical facts regarding divergences and singularities.
There are two basic types of divergence that can happen in a Fourier
series, divergence to infinity and indefinite oscillations or endless
wandering. If the series diverges everywhere over its periodic domain,
then the divergences may occur for two reasons, either there may exist no
real function that gives the Fourier coefficients of that series, or there
may be a failure of the internal mathematical machinery to represent
correctly an existing real function. On the other hand, if there is
convergence almost everywhere, and only one or more isolated points of
divergence to infinity, then it is likely that the divergences are caused
by the real function actually having integrable singularities at these
isolated points.

Only very radical divergence at all points within the domain can possibly
imply the actual non-existence of a real function that gives the
coefficients of the series. This is discussed in~\cite{FTotCPI}
and~\cite{FTotCPII}, in terms of an analytic structure that leads to a
simple and natural classification of divergences and singularities. The
typical case would be that in which the coefficients of a trigonometric
series diverge exponentially with $k$ when $k\to\infty$, in which case the
trigonometric series may fail to be a Fourier series at all. This is
seldom the case, so that in general we have either oscillatory divergence
almost everywhere, signifying a failure of the internal mathematical
machinery to represent faithfully an existing real function, or divergence
to infinity at isolated points where the real function being represented
by the Fourier series has actual integrable singularities.

In strict physical terms every divergence represents a failure to
represent or describe the physical world adequately. This means that
either the fundamental physical theory being used has failed, or that the
mathematical representation of the particular physical system at hand is
inadequate. The latter is much more often the case than the former, with
the description of the system being usually either oversimplified or
incomplete. For well-established fundamental physical theories being used
in a well-established domain of validity, the possibility of a fundamental
failure of the theory is an extremely remote one. On the other hand,
oversimplification of initial or boundary conditions is a relatively
common occurrence.

It is often possible to greatly simplify the use of the filters, avoiding
the necessity to solve the boundary value problem all over again after the
application of the filter. This is a consequence of the fact that the
filter operation often commutes with the differential operator contained
within the partial differential equation. In order to see this, let us
recall that the first-order filter can be understood as an integral
operator, which acts on the space of integrable real functions, since it
maps each real function to another real function. Let us show that the
elements of the Fourier basis are eigenfunctions of this operator. If we
apply the filter as defined in Equation~(\ref{deffilt}) to one of the
cosine functions of the basis we get

\noindent
\begin{eqnarray*}
  \frac{1}{2\varepsilon}
  \int_{x-\varepsilon}^{x+\varepsilon}dx'\,
  \cos\!\left(kx'\right)
  & = &
  \frac{1}{2\varepsilon k}
  \sin\!\left(kx'\right)\at{x-\varepsilon}{x+\varepsilon}
  \\
  & = &
  \frac{1}{2\varepsilon k}
  \left[
    \sin(kx+k\varepsilon)
    -
    \sin(kx-k\varepsilon)
  \right]
  \\
  & = &
  \frac{1}{2\varepsilon k}
  \left[
    \sin(kx)
    \cos(k\varepsilon)
    +
    \cos(kx)
    \sin(k\varepsilon)
  \right.
  \\
  &   &
  \hspace{2em}
  -
  \left.
    \sin(kx)
    \cos(k\varepsilon)
    +
    \cos(kx)
    \sin(k\varepsilon)
  \right]
  \\
  & = &
  \left[
    \frac{\sin(k\varepsilon)}{(k\varepsilon)}
  \right]
  \cos(kx).
\end{eqnarray*}

\noindent
This establishes the result, and also determines the eigenvalue, given by
the ratio shown within brackets. Once again we see here the sinc function
of the variable $(k\varepsilon)$, the same that appears in the Fourier
expansions of the kernels. The same can be done for the sine functions,
yielding

\noindent
\begin{eqnarray*}
  \frac{1}{2\varepsilon}
  \int_{x-\varepsilon}^{x+\varepsilon}dx'\,
  \sin\!\left(kx'\right)
  & = &
  -\,
  \frac{1}{2\varepsilon k}
  \cos\!\left(kx'\right)\at{x-\varepsilon}{x+\varepsilon}
  \\
  & = &
  -\,
  \frac{1}{2\varepsilon k}
  \left[
    \cos(kx+k\varepsilon)
    -
    \cos(kx-k\varepsilon)
  \right]
  \\
  & = &
  -\,
  \frac{1}{2\varepsilon k}
  \left[
    \cos(kx)
    \cos(k\varepsilon)
    -
    \sin(kx)
    \sin(k\varepsilon)
  \right.
  \\
  &   &
  \hspace{3em}
  -
  \left.
    \cos(kx)
    \cos(k\varepsilon)
    -
    \sin(kx)
    \sin(k\varepsilon)
  \right]
  \\
  & = &
  \left[
    \frac{\sin(k\varepsilon)}{(k\varepsilon)}
  \right]
  \sin(kx).
\end{eqnarray*}

\noindent
Note that this establishes the fact that these are the eigenfunctions of
the filter operator for all values of $\varepsilon$ in $(0,\pi)$. In other
words, this fact is stable by small variations of the real parameter
$\varepsilon$. As one can see, we have here the same eigenvalue as in the
previous case. There is therefore a degenerescence between each pair of
elements of the basis with the same value of $k$. It is also possible to
show that, up to this degenerescence, and assuming the stability by small
changes of $\varepsilon$, the elements of the Fourier basis are the {\em
  only} eigenfunctions of the filter operator when defined within the
periodic interval, as one can see in Section~\ref{APPeigenfunc} of
Appendix~\ref{APPpropfilt}. What all this means is that the filter acts in
an extremely simple way on the Fourier expansions. If we have the Fourier
expansion of the real function $f(x)$ in the periodic interval
$[-\pi,\pi]$,

\begin{displaymath}
  f(x)
  =
  \frac{1}{2}\,
  \alpha_{0}
  +
  \sum_{k=1}^{\infty}
  \left[
    \alpha_{k}\cos(kx)
    +
    \beta_{k}\sin(kx)
  \right],
\end{displaymath}

\noindent

\noindent
it follows at once that the corresponding expansion for the filtered
function is

\begin{displaymath}
  f_{\varepsilon}(x)
  =
  \frac{1}{2}\,
  \alpha_{0}
  +
  \sum_{k=1}^{\infty}
  \left\{
    \alpha_{k}
    \left[
      \frac{\sin(k\varepsilon)}{(k\varepsilon)}
    \right]
    \cos(kx)
    +
    \beta_{k}
    \left[
      \frac{\sin(k\varepsilon)}{(k\varepsilon)}
    \right]
    \sin(kx)
  \right\}.
\end{displaymath}

\noindent
What this means is that the Fourier coefficients $\alpha_{\varepsilon,k}$
and $\beta_{\varepsilon,k}$ of $f_{\varepsilon}(x)$ are given by

\noindent
\begin{eqnarray*}
  \alpha_{\varepsilon,k}
  & = &
  \left[
    \frac{\sin(k\varepsilon)}{(k\varepsilon)}
  \right]
  \alpha_{k},
  \\
  \beta_{\varepsilon,k}
  & = &
  \left[
    \frac{\sin(k\varepsilon)}{(k\varepsilon)}
  \right]
  \beta_{k},
\end{eqnarray*}

\noindent
a fact that can be shown directly and independently of the operator-based
argument used here, as one can see in Section~\ref{APPfourcoefs} of
Appendix~\ref{APPpropfilt}. Since the $\sin(k\varepsilon)$ in the
numerator of the ratio within brackets is a limited function, while in the
denominator we have simply $(k\varepsilon)$, in terms of the asymptotic
behavior of the coefficients the inclusion of the ratio, and hence the
action of the filter, corresponds simply to the inclusion of a factor of
$k$ in the denominator.

Since the elements of the Fourier basis are also eigenfunctions of the
second-derivative operator, as one can easily see by simply calculating
the derivatives,

\noindent
\begin{eqnarray*}
  \frac{\partial^{2}}{\partial x^{2}}\cos(kx)
  & = &
  -k^{2}\cos(kx),
  \\
  \frac{\partial^{2}}{\partial x^{2}}\sin(kx)
  & = &
  -k^{2}\sin(kx),
\end{eqnarray*}

\noindent
we may conclude that within the periodic interval the second-derivative
operator and the first-order low-pass filter operator have a complete set
of functions as a common set of eigenfunctions. It follows that the two
operators commute, a result which can be immediately extended to the
higher-order filters. Therefore, given any partial differential equation
which is purely second-order on the variable $x$ on which the filter acts,
and whose coefficients do not depend on that variable, it follows that if
a function $f(x)$ solves the equation, then the filtered function
$f_{\varepsilon}(x)$ is also a solution.

This leads to the fact that one may apply the filter directly to the
solution of the unfiltered problem, thus obtaining the same result that
one would obtain by first applying the filter to the initial or boundary
conditions and then solving the boundary value problem all over again.
This is the case for the Laplace equation, the wave equation and the
diffusion equation, in either Cartesian or cylindrical coordinates. Since
the unfiltered solution is represented in terms of a (possibly divergent)
Fourier series, in such circumstances it is immediate to write down the
filtered solution, by simply plugging the filter factor given by the sinc
function into the coefficients of the Fourier series obtained in the usual
way, as is illustrated by the examples in Appendix~\ref{APPexampfilt}.
Since it usually takes much more work to solve the boundary value problem
with the filtered initial or boundary conditions than to solve the
corresponding unfiltered problem, this can save a lot of work and effort.

\section{Conclusions}

Linear low-pass filters of arbitrary orders can be easily defined on the
real line, in a very simple way, either on the whole line or within a
periodic interval. We presented a definition of such filters in precise
mathematical terms, and also wrote them as linear integral operators
acting in the space of integrable real functions, expressed as integrals
involving certain kernel functions. We established several of the main
properties of the first-order filter. Due to the linearity of the filters
some of these properties, those involving the concept of invariance, are
immediately generalizable to the higher-order filters.

The use of the filters on divergent Fourier series produces other series
which are convergent, but which remain closely related to the original
problem within the physics application being dealt with, so long as the
range $\varepsilon$ is sufficiently small. It also produces series that
converge faster to smoother functions, and that can be differentiated a
certain number of times, as required by the applications involved, without
resulting in divergent series. We thus acquire a useful set of tools to
deal with Fourier series in a way that has a clear physical meaning in the
context of applications in physics. This set of tools can then be used as
a probe into the physical structure of the problems being dealt with.

It also follows in a very simple way that the filter operators commute
with the second derivative operator. We showed this in the case of the
first-order filter, and since the higher-order filters are just the
first-order one applied successive times, the result is immediately
extended to them. This leads to the fact that in many cases one may obtain
the solutions of the filtered boundary valuer problems by the simple
application of the filters directly to the solutions of the unfiltered
problems. This is very easily done when the solutions are expressed as
Fourier series, and may greatly simplify matters in practice.

Since any limited integrable function will have a limited set of Fourier
coefficients, it suffices to apply to such functions the first-order
linear low-pass filter of range $\varepsilon$ twice, or the second-order
filter of range $\varepsilon$ just once, in order to ensure the absolute
and uniform convergence of the resulting Fourier series. Therefore, it
becomes clear that any divergence of the original series must be due to
detailed structure that exists below the length scale that characterizes
these filters. For small enough values of the range of the filters such
structure cannot have a bearing on the physics involved, which allows us
to use the filters in this way. But besides this practical application the
filters give us a simple, clear and intuitive way to understand the origin
of the eventual divergences of the Fourier series.

\section{Acknowledgements}

The author would like to thank his friend and colleague Prof. Carlos
Eugênio Imbassay Carneiro, to whom he is deeply indebted for all his
interest and help, as well as his careful reading of the manuscript and
helpful criticism regarding this work.

\appendix

\section{Appendix: Properties of the First-Order
  Filter}\label{APPpropfilt}

Here we present simple proofs of the main properties of the first-order
linear low-pass filter on the real line of the coordinate $x$.

\subsection{Invariance of Linear Functions}

Let us show that if $f(x)$ is a linear function on the real line, then
$f_{\varepsilon}(x)=f(x)$. It suffices to simply calculate
$f_{\varepsilon}(x)$. We have $f(x)=a+bx$, so that

\noindent
\begin{eqnarray*}
  f_{\varepsilon}(x)
  & = &
  \frac{1}{2\varepsilon}
  \int_{x-\varepsilon}^{x+\varepsilon}dx'\,
  \left(a+bx'\right)
  \\
  & = &
  \frac{1}{2\varepsilon}
  \left(ax'+b\,\frac{x^{\prime 2}}{2}\right)
  \at{x-\varepsilon}{x+\varepsilon}
  \\
  & = &
  \frac{1}{2\varepsilon}
  \left[
    a
    \left(
      x+\varepsilon
      -
      x+\varepsilon
    \right)
    +
    b
    \left(
      \frac{x^{2}+2x\varepsilon+\varepsilon^{2}}{2}
      -
      \frac{x^{2}-2x\varepsilon+\varepsilon^{2}}{2}
    \right)
  \right]
  \\
  & = &
  \frac{1}{2\varepsilon}
  \left(
    2a\varepsilon
    +
    2bx\varepsilon
  \right)
  \\
  & = &
  a
  +
  2bx
  \\
  & = &
  f(x).
\end{eqnarray*}

\noindent
Note that if a function is defined in a piece-wise fashion, in any section
where it is linear the filter is the identity at all points $x$ where the
interval $(x-\varepsilon,x+\varepsilon)$ fits completely inside the
section. Therefore, in the $\varepsilon\to 0$ limit the filter becomes the
identity in the whole interior of such a section.

\subsection{Action on Powers and Polynomials}

Let us determine the action of the filter on a function which is a simple
power on the real line. If $f(x)=x^{n}$ then we have

\noindent
\begin{eqnarray*}
  f_{\varepsilon}(x)
  & = &
  \frac{1}{2\varepsilon}
  \int_{x-\varepsilon}^{x+\varepsilon}dx\,
  x^{\prime n}
  \\
  & = &
  \frac{1}{2\varepsilon(n+1)}\,
  x^{\prime(n+1)}\at{x-\varepsilon}{x+\varepsilon}
  \\
  & = &
  \frac{1}{2\varepsilon(n+1)}
  \left[
    (x+\varepsilon)^{n+1}
    -
    (x-\varepsilon)^{n+1}
  \right]
  \\
  & = &
  \frac{1}{2\varepsilon(n+1)}
  \sum_{k=0}^{n+1}
  \frac{(n+1)!}{k!(n+1-k)!}\,
  x^{n+1-k}\varepsilon^{k}
  \left[1-(-1)^{k}\right]
  \\
  & = &
  \frac{1}{\varepsilon(n+1)}
  \sum_{j=0}^{j_{M}}
  \frac{(n+1)!}{k!(n+1-k)!}\,
  x^{n+1-k}\varepsilon^{k},
\end{eqnarray*}

\noindent
where $k=2j+1$ and $j_{M}=n/2$ if $n$ is even, while $j_{M}=(n-1)/2$ if
$n$ is odd. We have therefore

\noindent
\begin{eqnarray*}
  f_{\varepsilon}(x)
  & = &
  \sum_{j=0}^{j_{M}}
  \frac{n!\varepsilon^{2j}}{(2j+1)!(n-2j)!}\,
  x^{n-2j}
  \\
  & = &
  x^{n}
  +
  \frac{n(n-1)\varepsilon^{2}}{3!}\,
  x^{n-2}
  +
  \frac{n(n-1)(n-2)(n-3)\varepsilon^{4}}{5!}\,
  x^{n-4}
  +
  \ldots\;.
\end{eqnarray*}

\noindent
We see therefore that the filter preserves the original power, and that
all other terms generated are of lower order and are damped by factors of
$\varepsilon^{2}$. It follows that the filter will reproduce any order-$n$
polynomial, adding to it a lower-order polynomial, of order $n-2$, with
all coefficients damped by powers of $\varepsilon^{2}$. Therefore, in the
$\varepsilon\to 0$ limit the filter reduces to the identity, in so far as
polynomials are concerned.

\subsection{Differentiability of Filtered Functions}\label{APPdifffilt}

Let us show that for any continuous function $f(x)$ the filtered function
$f_{\varepsilon}(x)$ is differentiable. We simply calculate the filtered
function $f_{\varepsilon}(x)$ at $x$ and $x+\Delta x$, then calculate its
variation $\Delta f_{\varepsilon}(x)$, divide by $\Delta x$ and finally
make $\Delta x\to 0$. The finite-difference ratio is given by

\noindent
\begin{eqnarray*}
  \frac{\Delta f_{\varepsilon}(x)}{\Delta x}
  & = &
  \frac{f_{\varepsilon}(x+\Delta x)-f_{\varepsilon}(x)}
  {\Delta x}
  \\
  & = &
  \frac{1}{2\varepsilon\Delta x}
  \int_{x+\Delta x-\varepsilon}^{x+\Delta x+\varepsilon}dx'\,
  f\!\left(x'\right)
  -
  \frac{1}{2\varepsilon\Delta x}
  \int_{x-\varepsilon}^{x+\varepsilon}dx'\,
  f\!\left(x'\right).
\end{eqnarray*}

\noindent
For any given value of $\varepsilon$, in the $\Delta x\to 0$ limit we will
eventually have $\Delta x\ll \varepsilon$, and then the domains of the two
integrals overlap in most of their extent, which we can see decomposing
the integrals as

\noindent
\begin{eqnarray*}
  \frac{\Delta f_{\varepsilon}(x)}{\Delta x}
  & = &
  \frac{1}{2\varepsilon\Delta x}
  \int_{x+\Delta x-\varepsilon}^{x+\varepsilon}dx'\,
  f\!\left(x'\right)
  +
  \frac{1}{2\varepsilon\Delta x}
  \int_{x+\varepsilon}^{x+\Delta x+\varepsilon}dx'\,
  f\!\left(x'\right)
  +
  \\
  &   &
  -\,
  \frac{1}{2\varepsilon\Delta x}
  \int_{x-\varepsilon}^{x+\Delta x-\varepsilon}dx'\,
  f\!\left(x'\right)
  -
  \frac{1}{2\varepsilon\Delta x}
  \int_{x+\Delta x-\varepsilon}^{x+\varepsilon}dx'\,
  f\!\left(x'\right)
  \\
  & = &
  \frac{1}{2\varepsilon}
  \left[
    \frac{1}{\Delta x}
    \int_{(x+\varepsilon)}^{(x+\varepsilon)+\Delta x}dx'\,
    f\!\left(x'\right)
  \right]
  -
  \frac{1}{2\varepsilon}
  \left[
    \frac{1}{\Delta x}
    \int_{(x-\varepsilon)}^{(x-\varepsilon)+\Delta x}dx'\,
    f\!\left(x'\right)
  \right].
\end{eqnarray*}

\noindent
We have here two integrals over intervals of length $\Delta x$, divided by
$\Delta x$. These normalized integrals give therefore the average values
of $f(x)$ around the points $x+\varepsilon$ and $x-\varepsilon$. Since the
function $f(x)$ is integrable these average values are finite, and since
it is continuous, the average value tends to the value of the function
when $\Delta x\to 0$, so that we get

\begin{displaymath}
  \lim_{\Delta x\to 0}
  \frac{\Delta f_{\varepsilon}(x)}{\Delta x}
  =
  \frac
  { 
    f\!\left(x+\varepsilon\right)
    -
    f\!\left(x-\varepsilon\right)
  }
  {2\varepsilon}.
\end{displaymath}

\noindent
This is true both for positive and negative values of $\Delta x$, and the
limit manifestly exists and has the value shown, which is independent of
the sign of $\Delta x$. Therefore, this establishes that
$f_{\varepsilon}(x)$ is differentiable.

\subsection{Continuity of Filtered Functions} 

Let us show that for any integrable function $f(x)$ the filtered function
$f_{\varepsilon}(x)$ is continuous. We simply calculate the filtered
function at $x$ and $x+\Delta x$ and then make $\Delta x\to 0$. The
variation of $f_{\varepsilon}(x)$ is given by

\noindent
\begin{eqnarray*}
  \Delta f_{\varepsilon}(x)
  & = &
  f_{\varepsilon}(x+\Delta x)-f_{\varepsilon}(x)
  \\
  & = &
  \frac{1}{2\varepsilon}
  \int_{x+\Delta x-\varepsilon}^{x+\Delta x+\varepsilon}dx'\,
  f\!\left(x'\right)
  -
  \frac{1}{2\varepsilon}
  \int_{x-\varepsilon}^{x+\varepsilon}dx'\,
  f\!\left(x'\right).
\end{eqnarray*}

\noindent
For any given value of $\varepsilon$, in the $\Delta x\to 0$ limit we will
eventually have $\Delta x\ll \varepsilon$, and then the domains of the two
integrals overlap in most of their extent, which we can see decomposing
the integrals as

\noindent
\begin{eqnarray*}
  \Delta f_{\varepsilon}(x)
  & = &
  \frac{1}{2\varepsilon}
  \int_{x+\Delta x-\varepsilon}^{x+\varepsilon}dx'\,
  f\!\left(x'\right)
  +
  \frac{1}{2\varepsilon}
  \int_{x+\varepsilon}^{x+\Delta x+\varepsilon}dx'\,
  f\!\left(x'\right)
  +
  \\
  &   &
  -\,
  \frac{1}{2\varepsilon}
  \int_{x-\varepsilon}^{x+\Delta x-\varepsilon}dx'\,
  f\!\left(x'\right)
  -
  \frac{1}{2\varepsilon}
  \int_{x+\Delta x-\varepsilon}^{x+\varepsilon}dx'\,
  f\!\left(x'\right)
  \\
  & = &
  \frac{1}{2\varepsilon}
  \left[
    \int_{(x+\varepsilon)}^{(x+\varepsilon)+\Delta x}dx'\,
    f\!\left(x'\right)
  \right]
  -
  \frac{1}{2\varepsilon}
  \left[
    \int_{(x-\varepsilon)}^{(x-\varepsilon)+\Delta x}dx'\,
    f\!\left(x'\right)
  \right].
\end{eqnarray*}

\noindent
We have here two integrals over intervals of length $\Delta x$. In the
$\Delta x\to 0$ limit we have integrals over zero-measure domains, and
since the function $f(x)$ is integrable, the result is zero,

\begin{displaymath}
  \lim_{\Delta x\to 0}
  \Delta f_{\varepsilon}(x)
  =
  0,
\end{displaymath}

\noindent
regardless of the sign of $\Delta x$, which establishes that
$f_{\varepsilon}(x)$ is continuous.

\subsection{Action on Dirac's Delta
  ``Function''}\label{APPactdelta} 

Let us assume that we have the ``function'' $f(x)=\delta(x-x_{0})$. Since
this is an integrable object, we may calculate the corresponding filtered
function, which as we shall see is in fact an actual function. The
function $f_{\varepsilon}(x)$ that corresponds to $f(x)$ through the
first-order filter of range $\varepsilon$ is, by definition,
    
\begin{displaymath}
  f_{\varepsilon}(x)
  =
  \frac{1}{2\varepsilon}
  \int_{x-\varepsilon}^{x+\varepsilon}dx'\,\delta(x'-x_{0}).
\end{displaymath}
  
\noindent
By the properties of the delta ``function'', this integral will be equal
to $1$ if the point $x_{0}$ is within the integration interval, and $0$ if
it is outside. The point $x_{0}$ can only be within the integration
interval if the distance between $x$ and $x_{0}$ is smaller than
$\varepsilon$, that is, if $|x-x_{0}|<\varepsilon$. Therefore we have for
the resulting function the piece-wise description
  
\noindent
\begin{displaymath}
  %
  \renewcommand{\arraystretch}{2.1}
  \begin{array}{rclcl}
    f_{\varepsilon}(x)
    & = &
    0
    & \mbox{if}
    & x<(x_{0}-\varepsilon),
    \\
    f_{\varepsilon}(x)
    & = &
    \FFrac{1}{2\varepsilon}
    & \mbox{if}
    & (x_{0}-\varepsilon)<x<(x_{0}+\varepsilon),
    \\
    f_{\varepsilon}(x)
    & = &
    0
    & \mbox{if}
    & (x_{0}+\varepsilon)<x.
  \end{array}
\end{displaymath}

\noindent
This is a rectangular pulse centered at $x_{0}$, with height
$1/(2\varepsilon)$ and width $(2\varepsilon)$, having therefore unit area.
Note that this is, in fact, the first-order kernel itself, that is

\begin{displaymath}
  f_{\varepsilon}(x)
  =
  K_{\varepsilon}^{(1)}(x-x_{0}).
\end{displaymath}

\noindent
This one-parameter family of functions is one that is commonly used for
the very definition the Dirac delta ``function'' in the limit
$\varepsilon\to 0$, and therefore we have that

\begin{displaymath}
  \lim_{\varepsilon\to 0}K_{\varepsilon}^{(1)}(x-x_{0})
  =
  \delta(x-x_{0}).
\end{displaymath}

\noindent
Looking at the filter as an operator in some larger space of integrable
objects, this means that it becomes the identity in the $\varepsilon\to 0$
limit, in so far as delta ``functions'' are concerned. Note that the delta
``function'' can also be understood as the kernel of an order-zero filter,

\begin{displaymath}
  f_{0}^{(0)}(x)
  =
  \int_{-\infty}^{\infty}dx'\,
  \delta\!\left(x-x'\right)f\!\left(x'\right).
\end{displaymath}

\noindent
This filter is the identity where $f(x)$ is continuous, so that typically
it is the identity almost everywhere. Note also that, as a particular case
of this expression, we may conclude that the first-order kernel is the
result of the application of the first-order filter to the delta
``functions'',

\begin{displaymath}
  K_{\varepsilon}^{(1)}\!\left(x-x''\right)
  =
  \int_{-\infty}^{\infty}dx'\,
  K_{\varepsilon}^{(1)}\!\left(x'-x''\right)
  \delta\!\left(x-x'\right),
\end{displaymath}

\noindent
which holds everywhere so long as the first-order kernel is defined as we
did in Equation~(\ref{defK1full}) and so long as we use the average of the
two lateral limits as the value given by the integral of the delta
``function'' at a point of discontinuity of the function involved.

\subsection{Reduction to the Identity}

Let us show that in the $\varepsilon\to 0$ limit the filter reduces to an
almost-identity operation, in the sense that it reproduces in the output
function $f_{\varepsilon}(x)$ the input function $f(x)$ almost everywhere.
If we consider the well-known relation mentioned in the previous section
as a possible definition of the Dirac delta ``function'', as the
$\varepsilon\to 0$ limit of the first-order kernel
$K_{\varepsilon}^{(1)}\!\left(x-x'\right)$, it becomes clear that we have,
for an arbitrary integrable function $f(x)$

\noindent
\begin{eqnarray*}
  \lim_{\varepsilon\to 0}
  f_{\varepsilon}(x)
  & = &
  \lim_{\varepsilon\to 0}
  \int_{-\infty}^{\infty}dx'\,
  K_{\varepsilon}^{(1)}\!\left(x-x'\right)
  f\!\left(x'\right)
  \\
  & = &
  \int_{-\infty}^{\infty}dx'\,
  \left[
    \,
    \lim_{\varepsilon\to 0}
    K_{\varepsilon}^{(1)}\!\left(x-x'\right)
  \right]
  f\!\left(x'\right)
  \\
  & = &
  \int_{-\infty}^{\infty}dx'\,
  \delta(x-x')
  f\!\left(x'\right).
\end{eqnarray*}

\noindent
According to the properties of the delta ``function'', this integral
returns the value $f(x)$ at every point where this function is continuous.
We therefore have

\begin{displaymath}
  \lim_{\varepsilon\to 0}
  f_{\varepsilon}(x)
  =
  f(x),
\end{displaymath}

\noindent
at every point where $f(x)$ is continuous. Since this may fail at a finite
(or at least zero-measure) set of points where $f(x)$ is discontinuous, we
say that in the $\varepsilon\to 0$ limit the first-order filter reduces to
the identity almost everywhere. We may also say that the filter becomes an
almost-identity operation in the limit. What happens at the points of
discontinuity of $f(x)$ is discussed in the next section.

\subsection{Points of Discontinuity} 

Let us show that in the $\varepsilon\to 0$ limit the function
$f_{\varepsilon}(x)$ essentially reproduces the original function
$f(x)$. Stating it more precisely, we will show that, if the function
$f(x)$ has an isolated point of discontinuity at $x_{0}$, then in the
$\varepsilon\to 0$ limit $f_{\varepsilon}(x_{0})$ tends to the average of
the two lateral limits of $f(x)$ to the point $x_{0}$, that is,

\begin{displaymath}
  \lim_{\varepsilon\to 0}f_{\varepsilon}(x_{0})
  =
  \frac{1}{2}\left({\cal L}_{+}+{\cal L}_{-}\right),
\end{displaymath}

\noindent
where

\begin{displaymath}
  {\cal L}_{\pm}
  =
  \lim_{x\to x_{0\pm}}f(x),
\end{displaymath}

\noindent
regardless of the value that $f(x)$ assumes at $x_{0}$. In particular, if
$f(x)$ is continuous at $x_{0}$, then ${\cal L}_{+}={\cal L}_{-}=f(x_{0})$
and hence $f_{\varepsilon}(x_{0})$ tends to $f(x_{0})$ in the limit, thus
reproducing the original function at that point.

Here is the proof: if $f(x)$ has an isolated point of discontinuity at
$x_{0}$, then there are two neighborhoods of $x_{0}$, one to the left and
another one to the right, where $f(x)$ is continuous. For sufficiently
small $\varepsilon$, the interval of integration will fit into this
combined neighborhood, so that the only point of discontinuity within it
will be $x_{0}$. Let us consider then the value of
$f_{\varepsilon}(x_{0})$, as given by the definition,

\begin{displaymath}
  f_{\varepsilon}(x_{0})
  =
  \frac{1}{2\varepsilon}
  \int_{x_{0}-\varepsilon}^{x_{0}+\varepsilon}dx'\,f(x').
\end{displaymath}
    
\noindent
We may separate this integral in two, one in the left neighborhood and
another one in the right neighborhood,
    
\begin{displaymath}
  f_{\varepsilon}(x_{0})
  =
  \frac{1}{2\varepsilon}
  \int_{x_{0}-\varepsilon}^{x_{0}}dx'\,f(x')
  +
  \frac{1}{2\varepsilon}
  \int_{x_{0}}^{x_{0}+\varepsilon}dx'\,f(x').
\end{displaymath}

\noindent
Since the function $f(x)$ is integrable, the integrals converge to
$\varepsilon$ times the average value of the function over each
sub-interval, so that we have

\begin{displaymath}
  \lim_{\varepsilon\to 0}f_{\varepsilon}(x_{0})
  =
  \frac{1}{2}
  \bar{f}(x_{0-})
  +
  \frac{1}{2}
  \bar{f}(x_{0+}).
\end{displaymath}

\noindent
Finally, since the function $f(x)$ is continuous in each sub-interval, in
the $\varepsilon\to 0$ limit each average value converges to the
corresponding lateral limit of $f(x)$, so that we have

\begin{displaymath}
  \lim_{\varepsilon\to 0}f_{\varepsilon}(x_{0})
  =
  \frac{1}{2}\left({\cal L}_{+}+{\cal L}_{-}\right),
\end{displaymath}

\noindent
where

\begin{displaymath}
  {\cal L}_{\pm}
  =
  \lim_{x\to x_{0\pm}}f(x).
\end{displaymath}

\noindent
This establishes the result. As a consequence of this, if $f(x)$ is
continuous at $x_{0}$, then ${\cal L}_{+}={\cal L}_{-}=f(x_{0})$, and
therefore we have

\begin{displaymath}
  \lim_{\varepsilon\to 0}f_{\varepsilon}(x_{0})
  =
  f(x_{0}).
\end{displaymath}

\noindent
We therefore conclude that in the $\varepsilon\to 0$ limit the filtered
function $f_{\varepsilon}(x)$ reproduces the original function $f(x)$
where it is continuous. There may be isolated points of discontinuity
where this fails, and therefore we say that in the $\varepsilon\to 0$
limit the filtered function $f_{\varepsilon}(x)$ reproduces the original
function $f(x)$ almost everywhere.

\subsection{Points of Non-Differentiability} 

Let us show that in the $\varepsilon\to 0$ limit the derivative of the
function $f_{\varepsilon}(x)$ essentially reproduces the derivative of the
original function $f(x)$. Stating it more precisely, we will show that, if
the function $f(x)$ is continuous but has an isolated point of
non-differentiability at $x_{0}$, then in the $\varepsilon\to 0$ limit the
derivative of $f_{\varepsilon}(x_{0})$ tends to the average of the two
lateral limits of the derivative of $f(x)$ to the point $x_{0}$, that is,

\begin{displaymath}
  \lim_{\varepsilon\to 0}\frac{df_{\varepsilon}}{dx}(x_{0})
  =
  \frac{1}{2}\left({\cal L'}_{+}+{\cal L'}_{-}\right),
\end{displaymath}
    
\noindent
where
    
\begin{displaymath}
  {\cal L'}_{\pm}
  =
  \lim_{x\to x_{0\pm}}\frac{df}{dx}(x),
\end{displaymath}

\noindent
regardless of any value that may be artificially given to the derivative
of $f(x)$ at $x_{0}$. In particular, if $f(x)$ is differentiable at
$x_{0}$, then ${\cal L'}_{+}$ and ${\cal L'}_{-}$ are both equal to the
derivative of $f(x)$ at $x_{0}$, and hence the derivative of
$f_{\varepsilon}(x)$ tends to the derivative of $f(x)$ at $x_{0}$ in the
limit, thus reproducing the derivative of the original function at that
point.

Here is the proof: if $f(x)$ is continuous at $x_{0}$ and has an isolated
point of discontinuity there, then there are two neighborhoods of $x_{0}$,
one to the left and another one to the right, where $f(x)$ is continuous
and differentiable. According to the results of Section~\ref{APPdifffilt}
of this Appendix, since $f(x)$ is continuous at and around $x_{0}$,
$f_{\varepsilon}(x)$ is differentiable and its derivative at $x_{0}$ is
given by

\begin{displaymath}
  \frac{df_{\varepsilon}}{dx}(x_{0})
  =
  \frac
  { 
    f(x_{0}+\varepsilon)
    -
    f(x_{0}-\varepsilon)
  }
  {2\varepsilon}.
\end{displaymath}

\noindent
However, since $f(x)$ is not differentiable at $x_{0}$, the
$\varepsilon\to 0$ limit of the right-hand side of this equation does not
give us any definite results. We may however separate this expression in
two, each one making reference to only one of the two neighborhoods,

\noindent
\begin{eqnarray*}
  \frac{df_{\varepsilon}}{dx}(x_{0})
  & = &
  \frac
  { 
    f(x_{0}+\varepsilon)
    -
    f\!\left(x_{0}\right)
    +
    f\!\left(x_{0}\right)
    -
    f(x_{0}-\varepsilon)
  }
  {2\varepsilon}
  \\
  & = &
  \frac{1}{2}\,
  \frac
  { 
    f(x_{0}+\varepsilon)
    -
    f(x_{0})
  }
  {\varepsilon}
  +
  \frac{1}{2}\,
  \frac
  { 
    f(x_{0})
    -
    f(x_{0}-\varepsilon)
  }
  {\varepsilon}.
\end{eqnarray*}

\noindent
It is now clear that, since $f(x)$ is differentiable in the two lateral
neighborhoods, in the $\varepsilon\to 0$ limit the two terms in the
right-hand side of this equation converge respectively to the right and
left derivatives of $f(x)$ at $x_{0}$. We therefore have

\begin{displaymath}
  \lim_{\varepsilon\to 0}
  \frac{df_{\varepsilon}}{dx}(x_{0})
  =
  \frac{1}{2}\left({\cal L'}_{+}+{\cal L'}_{-}\right),
\end{displaymath}

\noindent
where
    
\begin{displaymath}
  {\cal L'}_{\pm}
  =
  \lim_{x\to x_{0\pm}}\frac{df}{dx}(x).
\end{displaymath}

\noindent
This establishes the result. In particular, if $f(x)$ is differentiable at
$x_{0}$, then ${\cal L'}_{+}$ and ${\cal L'}_{-}$ are both equal to the
derivative of $f(x)$ at $x_{0}$, and therefore we have for the derivative
of $f_{\varepsilon}(x)$ at $x_{0}$

\begin{displaymath}
  \lim_{\varepsilon\to 0}
  \frac{df_{\varepsilon}}{dx}(x_{0})
  =
  \frac{df}{dx}(x_{0}),
\end{displaymath}

\noindent
thus reproducing the derivative of the original function at that point.

\subsection{Invariance of Definite Integrals}\label{APPinvints}

Let us determine the effect of the filter on the definite integral of a
function $f(x)$ with compact support on the real line. We may write the
integral as

\begin{displaymath}
  I
  =
  \int_{-\infty}^{\infty}dx\,
  f(x),
\end{displaymath}

\noindent
where the integrand is non-zero only inside a closed interval. The
integral of the filtered function $f_{\varepsilon}(x)$ has support on
another closed interval, that of $f(x)$ increased by $\varepsilon$ in each
direction, and is similarly given by

\noindent
\begin{eqnarray*}
  I_{\varepsilon}
  & = &
  \int_{-\infty}^{\infty}dx\,
  f_{\varepsilon}(x)
  \\
  & = &
  \int_{-\infty}^{\infty}dx\,
  \frac{1}{2\varepsilon}
  \int_{x-\varepsilon}^{x+\varepsilon}dx'\,
  f\!\left(x'\right)
  \\
  & = &
  \frac{1}{2\varepsilon}
  \int_{-\infty}^{\infty}dx\,
  \int_{x-\varepsilon}^{x+\varepsilon}dx'\,
  f\!\left(x'\right),
\end{eqnarray*}

\noindent
where we used the definition of $f_{\varepsilon}(x)$ in terms of $f(x)$.
We now make the change of variables $x'=x''+x$ on the inner integral,
implying $dx'=dx''$ and $x''=x'-x$, leading to

\noindent
\begin{eqnarray*}
  I_{\varepsilon}
  & = &
  \frac{1}{2\varepsilon}
  \int_{-\infty}^{\infty}dx\,
  \int_{-\varepsilon}^{\varepsilon}dx''\,
  f(x''+x),
  \\
  & = &
  \frac{1}{2\varepsilon}
  \int_{-\varepsilon}^{\varepsilon}dx''\,
  \int_{-\infty}^{\infty}dx\,
  f(x''+x).
\end{eqnarray*}

\noindent
Since the integral on $x$ is over the whole real line, we may now change
variables on it without changing the integration limits, using $x=x'-x''$,
with $x'=x''+x$ and $dx=dx'$, and thus obtaining

\noindent
\begin{eqnarray*}
  I_{\varepsilon}
  & = &
  \frac{1}{2\varepsilon}
  \int_{-\varepsilon}^{\varepsilon}dx''\,
  \int_{-\infty}^{\infty}dx'\,
  f\!\left(x'\right)
  \\
  & = &
  \frac{1}{2\varepsilon}
  \int_{-\varepsilon}^{\varepsilon}dx''\,
  I
  \\
  & = &
  I,
\end{eqnarray*}

\noindent
were we recognized the form of the integral $I$. In this way we show that
$I_{\varepsilon}=I$, that is, the filter does not change the definite
integral at all. Another way to state this is to say that the filter does
not change the average value of $f(x)$ over the common support of $f(x)$
and $f_{\varepsilon}(x)$.

\subsection{Periodicity of Filtered Functions}

Let us show that if $f(x)$ is a periodic function, with a period that we
choose arbitrarily to be $2\pi$, then $f_{\varepsilon}(x)$ is also
periodic, with the same period. It suffices to simply calculate

\noindent
\begin{eqnarray*}
  f_{\varepsilon}(x+2\pi)
  & = &
  \frac{1}{2\varepsilon}
  \int_{x+2\pi-\varepsilon}^{x+2\pi+\varepsilon}dx'\,
  f\!\left(x'\right)
  \\
  & = &
  \frac{1}{2\varepsilon}
  \int_{x-\varepsilon}^{x+\varepsilon}dx''\,
  f\!\left(x''+2\pi\right),
\end{eqnarray*}

\noindent
where we changed variables to $x''=x'-2\pi$, so that $x'=x''+2\pi$. Since
$f(x)$ is periodic with period $2\pi$, we now have

\noindent
\begin{eqnarray*}
  f_{\varepsilon}(x+2\pi)
  & = &
  \frac{1}{2\varepsilon}
  \int_{x-\varepsilon}^{x+\varepsilon}dx''\,
  f\!\left(x''\right)
  \\
  & = &
  f_{\varepsilon}(x),
\end{eqnarray*}

\noindent
so that we may conclude that $f_{\varepsilon}(x)$ is periodic with period
$2\pi$.

\subsection{Invariance of Averages Over the Period}

Let us determine the effect of the filter on the Fourier coefficient
$\alpha_{0}$. We start with the coefficient of $f(x)$, which is given by

\begin{displaymath}
  \alpha_{0}
  =
  \frac{1}{\pi}
  \int_{-\pi}^{\pi}dx\,
  f(x),
\end{displaymath}

\noindent
where we arbitrarily chose $[-\pi,\pi]$ as the periodic interval. The
Fourier coefficient $\alpha_{\varepsilon,0}$ of $f_{\varepsilon}(x)$ is
similarly given by

\noindent
\begin{eqnarray*}
  \alpha_{\varepsilon,0}
  & = &
  \frac{1}{\pi}
  \int_{-\pi}^{\pi}dx\,
  f_{\varepsilon}(x)
  \\
  & = &
  \frac{1}{\pi}
  \int_{-\pi}^{\pi}dx\,
  \frac{1}{2\varepsilon}
  \int_{x-\varepsilon}^{x+\varepsilon}dx'\,
  f\!\left(x'\right)
  \\
  & = &
  \frac{1}{2\varepsilon\pi}
  \int_{-\pi}^{\pi}dx\,
  \int_{x-\varepsilon}^{x+\varepsilon}dx'\,
  f\!\left(x'\right),
\end{eqnarray*}

\noindent
where we used the definition of $f_{\varepsilon}(x)$ in terms of $f(x)$.
We now make the change of variables $x'=x''+x$ on the inner integral,
implying $dx'=dx''$ and $x''=x'-x$, leading to

\noindent
\begin{eqnarray*}
  \alpha_{\varepsilon,0}
  & = &
  \frac{1}{2\varepsilon\pi}
  \int_{-\pi}^{\pi}dx\,
  \int_{-\varepsilon}^{\varepsilon}dx''\,
  f(x''+x),
  \\
  & = &
  \frac{1}{2\varepsilon}
  \int_{-\varepsilon}^{\varepsilon}dx''\,
  \frac{1}{\pi}
  \int_{-\pi}^{\pi}dx\,
  f(x''+x).
\end{eqnarray*}

\noindent
Since the integral on $x$ is over the whole period, we may now change
variables on it without changing the integration limits, using $x=x'-x''$,
with $x'=x''+x$ and $dx=dx'$, and thus obtaining

\noindent
\begin{eqnarray*}
  \alpha_{\varepsilon,0}
  & = &
  \frac{1}{2\varepsilon}
  \int_{-\varepsilon}^{\varepsilon}dx''\,
  \frac{1}{\pi}
  \int_{-\pi}^{\pi}dx'\,
  f\!\left(x'\right)
  \\
  & = &
  \frac{1}{2\varepsilon}
  \int_{-\varepsilon}^{\varepsilon}dx''\,
  \alpha_{0}
  \\
  & = &
  \alpha_{0},
\end{eqnarray*}

\noindent
were we recognized the form of $\alpha_{0}$. In this way we show that
$\alpha_{\varepsilon,0}=\alpha_{0}$, that is, the filter does not change
$\alpha_{0}$ at all. Another way to state this is to say that the filter
does not change the average value of $f(x)$ over the periodic interval.

\subsection{Action on the Fourier Coefficients with {\boldmath
    $k>0$}}\label{APPfourcoefs}

Let us determine the effect of the filter on the Fourier coefficients for
$k>0$. We start with the Fourier coefficients of $f(x)$, which are given
by

\noindent
\begin{eqnarray*}
  \alpha_{k}
  & = &
  \frac{1}{\pi}
  \int_{-\pi}^{\pi}dx\,
  f(x)
  \cos(kx),
  \\
  \beta_{k}
  & = &
  \frac{1}{\pi}
  \int_{-\pi}^{\pi}dx\,
  f(x)
  \sin(kx),
\end{eqnarray*}

\noindent
where we arbitrarily chose $[-\pi,\pi]$ as the periodic interval. The
Fourier coefficients of $f_{\varepsilon}(x)$ are similarly given by

\noindent
\begin{eqnarray*}
  \alpha_{\varepsilon,k}
  & = &
  \frac{1}{\pi}
  \int_{-\pi}^{\pi}dx\,
  f_{\varepsilon}(x)
  \cos(kx),
  \\
  \beta_{\varepsilon,k}
  & = &
  \frac{1}{\pi}
  \int_{-\pi}^{\pi}dx\,
  f_{\varepsilon}(x)
  \sin(kx).
\end{eqnarray*}

\noindent
Let us work out only the first case, since the work for the second one in
essentially identical. Using the definition of $f_{\varepsilon}(x)$ in
terms of $f(x)$ we have

\noindent
\begin{eqnarray*}
  \alpha_{\varepsilon,k}
  & = &
  \frac{1}{\pi}
  \int_{-\pi}^{\pi}dx\,
  \cos(kx)\,
  \frac{1}{2\varepsilon}
  \int_{x-\varepsilon}^{x+\varepsilon}dx'\,
  f\!\left(x'\right)
  \\
  & = &
  -\,
  \frac{1}{2\varepsilon\pi}
  \int_{-\pi}^{\pi}dx\,
  \frac{\sin(kx)}{k}\,
  \frac{d}{dx}
  \int_{x-\varepsilon}^{x+\varepsilon}dx'\,
  f\!\left(x'\right)
  \\
  & = &
  -\,
  \frac{1}{2\varepsilon\pi k}
  \int_{-\pi}^{\pi}dx\,
  \sin(kx)\;
  f\!\left(x'\right)\at{x-\varepsilon}{x+\varepsilon}
  \\
  & = &
  -\,
  \frac{1}{2\varepsilon\pi k}
  \int_{-\pi}^{\pi}dx\,
  \sin(kx)
  f(x+\varepsilon)
  +
  \frac{1}{2\varepsilon\pi k}
  \int_{-\pi}^{\pi}dx\,
  \sin(kx)
  f(x-\varepsilon),
\end{eqnarray*}

\noindent
where we integrated by parts and where there is no integrated term due to
the periodicity of the integrand on the domain. We now change variables in
each integral, using $x'=x\pm\varepsilon$, in order to obtain

\noindent
\begin{eqnarray*}
  \alpha_{\varepsilon,k}
  & = &
  -\,
  \frac{1}{2\varepsilon\pi k}
  \int_{-\pi}^{\pi}dx'\,
  \sin(kx'-k\varepsilon)
  f\!\left(x'\right)
  +
  \frac{1}{2\varepsilon\pi k}
  \int_{-\pi}^{\pi}dx'\,
  \sin(kx'+k\varepsilon)
  f\!\left(x'\right)
  \\
  & = &
  \frac{1}{2\varepsilon\pi k}
  \int_{-\pi}^{\pi}dx'\,
  f\!\left(x'\right)
  \left[
    \sin(kx'+k\varepsilon)
    -
    \sin(kx'-k\varepsilon)
  \right],
\end{eqnarray*}

\noindent
where the integration limits did not change in the transformations of
variables due to the periodicity of the integrand on the domain. We are
left with

\noindent
\begin{eqnarray*}
  \alpha_{\varepsilon,k}
  & = &
  \frac{1}{2\varepsilon\pi k}
  \int_{-\pi}^{\pi}dx'\,
  f\!\left(x'\right)
  \left[
    \sin\!\left(kx'\right)
    \cos(k\varepsilon)
    +
    \sin(k\varepsilon)
    \cos\!\left(kx'\right)
  \right.
  +
  \\
  &   &
  \hspace{9em}
  -
  \left.
    \sin\!\left(kx'\right)
    \cos(k\varepsilon)
    +
    \sin(k\varepsilon)
    \cos\!\left(kx'\right)
  \right]
  \\
  & = &
  \left[
    \frac{\sin(k\varepsilon)}{(k\varepsilon)}
  \right]
  \frac{1}{\pi}
  \int_{-\pi}^{\pi}dx'\,
  f\!\left(x'\right)
  \cos\!\left(kx'\right).
\end{eqnarray*}

\noindent
Since we recover in this way the expression of the Fourier coefficients
$\alpha_{k}$ of $f(x)$, we get

\begin{displaymath}
  \alpha_{\varepsilon,k}
  =
  \left[
    \frac{\sin(k\varepsilon)}{(k\varepsilon)}
  \right]
  \alpha_{k},
\end{displaymath}

\noindent
and repeating the calculation for the other coefficients one gets

\begin{displaymath}
  \beta_{\varepsilon,k}
  =
  \left[
    \frac{\sin(k\varepsilon)}{(k\varepsilon)}
  \right]
  \beta_{k}.
\end{displaymath}

\noindent
Once again we see the sinc function of the variable $(k\varepsilon)$
appearing here. Since $\sin(k\varepsilon)$ is a limited function, in terms
of the asymptotic behavior of the coefficients, for large values of $k$,
the net effect of the filter is to add a factor of $k$ to the denominator.

\subsection{Completeness of the Set of
  Eigenfunctions}\label{APPeigenfunc} 

Let us show that, up to the degeneracy between the pairs of elements of
the basis with the same $k$, the elements of the Fourier basis are the
only eigenfunctions of the filter operator, when it is defined within the
periodic interval. In order to do this, let us first point out that two
eigenvalues, for two different values of $k$, are never equal. We can see
this assuming that there are positive values $k$ and $k'$ such that

\begin{displaymath}
  \frac{\sin(k\varepsilon)}{(k\varepsilon)}
  =
  \frac{\sin\!\left(k'\varepsilon\right)}{\left(k'\varepsilon\right)}.
\end{displaymath}

\noindent
Since this must stay valid for small changes of $\varepsilon$, we may
differentiate with respect to $\varepsilon$ and thus obtain

\begin{displaymath}
  \cos(k\varepsilon)
  =
  \cos\!\left(k'\varepsilon\right).
\end{displaymath}

\noindent
This now implies that

\begin{displaymath}
  \sin(k\varepsilon)
  =
  \pm\sin\!\left(k'\varepsilon\right).
\end{displaymath}

\noindent
Since $\varepsilon>0$ and both $k$ and $k'$ are positive, we must have

\begin{displaymath}
  \sin(k\varepsilon)
  =
  \sin\!\left(k'\varepsilon\right).
\end{displaymath}

\noindent
Since both the cosines and the sines of the two arguments are thus seen to
be equal, it follows that the two arguments must be equal, and hence that
we must have $k=k'$. Therefore, the eigenvalues for two different values
of $k$ are never equal. Let us consider now an arbitrary function $f(x)$
and its expression in the Fourier basis,

\begin{displaymath}
  f(x)
  =
  \frac{1}{2}\,
  \alpha_{0}
  +
  \sum_{k=1}^{\infty}
  \left[
    \alpha_{k}\cos(kx)
    +
    \beta_{k}\sin(kx)
  \right],
\end{displaymath}

\noindent
Let us assume that this function is not identically zero, and that it is
an eigenfunction of the first-order filter operator, that is

\begin{displaymath}
  \frac{1}{2\varepsilon}
  \int_{x-\varepsilon}^{x+\varepsilon}dx'\,
  f\!\left(x'\right)
  =
  \lambda f(x),
\end{displaymath}

\noindent
for some real number $\lambda$. We are therefore assuming that it is a
normalizable function that is an eigenfunction of the filter operator.
Using the expression of the function on the Fourier basis, which is
complete to represent almost everywhere any integrable real function on
the periodic interval, we get

\noindent
\begin{eqnarray*}
  \lefteqn
  {
    \frac{1}{2\varepsilon}
    \int_{x-\varepsilon}^{x+\varepsilon}dx'\,
    \left[
      \frac{1}{2}\,
      \alpha_{0}
      +
      \sum_{k=1}^{\infty}
      \alpha_{k}\cos\!\left(kx'\right)
      +
      \sum_{k=1}^{\infty}
      \beta_{k}\sin\!\left(kx'\right)
    \right]
  }
  \hspace{3em}
  &   &
  \\
  & = &
  \lambda
  \left[
    \frac{1}{2}\,
    \alpha_{0}
    +
    \sum_{k=1}^{\infty}
    \alpha_{k}\cos(kx)
    +
    \sum_{k=1}^{\infty}
    \beta_{k}\sin(kx)
  \right]
  \;\;\;\Rightarrow
  \\
  \lefteqn
  {
    \frac{1}{2}\,
    \alpha_{0}
    +
    \sum_{k=1}^{\infty}
    \left[
      \frac{\sin(k\varepsilon)}{(k\varepsilon)}
    \right]
    \alpha_{k}\cos(kx)
    +
    \sum_{k=1}^{\infty}
    \left[
      \frac{\sin(k\varepsilon)}{(k\varepsilon)}
    \right]
    \beta_{k}\sin(kx)
  }
  \hspace{3em}
  &   &
  \\
  & = &
  \frac{1}{2}\,
  \lambda
  \alpha_{0}
  +
  \sum_{k=1}^{\infty}
  \lambda
  \alpha_{k}\cos(kx)
  +
  \sum_{k=1}^{\infty}
  \lambda
  \beta_{k}\sin(kx).
\end{eqnarray*}

\noindent
Passing all terms to the same side we may write this as the expansion of a
certain function in the Fourier basis,

\begin{displaymath}
  \frac{1}{2}\,
  (1-\lambda)
  \alpha_{0}
  +
  \sum_{k=1}^{\infty}
  \left[
    \frac{\sin(k\varepsilon)}{(k\varepsilon)}
    -
    \lambda
  \right]
  \alpha_{k}\cos(kx)
  +
  \sum_{k=1}^{\infty}
  \left[
    \frac{\sin(k\varepsilon)}{(k\varepsilon)}
    -
    \lambda
  \right]
  \beta_{k}\sin(kx)
  =
  0.
\end{displaymath}

\noindent
This is the expansion of the null function in the Fourier basis, which is
unique and therefore implies that all coefficients must be zero. We have
therefore

\noindent
\begin{eqnarray*}
  (1-\lambda)
  \alpha_{0}
  & = &
  0,
  \\
  \left[
    \frac{\sin(k\varepsilon)}{(k\varepsilon)}
    -
    \lambda
  \right]
  \alpha_{k}
  & = &
  0,
  \\
  \left[
    \frac{\sin(k\varepsilon)}{(k\varepsilon)}
    -
    \lambda
  \right]
  \beta_{k}
  & = &
  0,
\end{eqnarray*}

\noindent
the last two for all $k>0$. Taking first the case $k=0$, if $\lambda=1$ we
may have $\alpha_{0}\neq 0$, but since the other eigenvalues are never
equal to $1$, we must have then $\alpha_{k}=0$ and $\beta_{k}=0$ for all
$k>0$. On the other hand, if $\lambda$ is equal to one of the eigenvalues
with $k>0$, then it is different from all the other eigenvalues, since the
eigenvalues for two values of $k$ are never equal. In this case we may
have $\alpha_{k}\neq 0$ and $\beta_{k}\neq 0$ for one value of $k$, but
all the other coefficients must be zero. Therefore the function $f(x)$
must be either a constant function or a linear combination of $\cos(kx)$
and $\sin(kx)$ for a single value of $k$.

\section{Appendix: Examples of Use of the First-Order
  Filter}\label{APPexampfilt}

In this appendix we will give a few illustrative examples of the use of
the first-order linear low-pass filter in physical systems, involving the
solution of boundary value problems of partial differential equations
through the use of Fourier series.

\subsection{The Plucked String}

\begin{figure}[ht]
  \centering
  \fbox{
    \epsfig{file=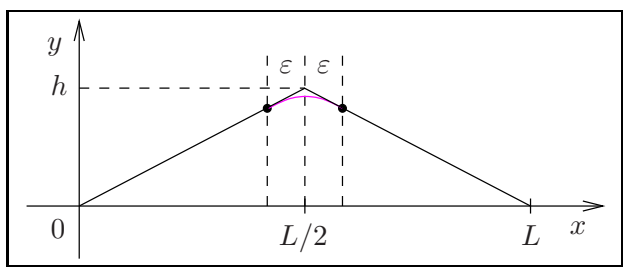,scale=1.0,angle=0}
  }
  \caption{Illustration of the original initial condition for the position
    of the plucked string, superposed with the filtered initial
    condition.}
  \label{Fig02}
\end{figure}

Consider the vibrating string of length $L$. In the small displacement
approximation its movement is given by the wave equation

\begin{displaymath}
  \frac{\partial^{2}f(x,t)}{\partial x^{2}}
  -
  \frac{1}{\nu^{2}}\,
  \frac{\partial^{2}f(x,t)}{\partial t^{2}}
  =
  0,
\end{displaymath}

\noindent
where $\nu$ is the speed of the waves on the string and where $f(x,t)$ is
the displacement from equilibrium at position $x$ and time $t$. The
boundary conditions are $f(0,t)=0$ and $f(L,t)=0$ for all $t$. Let us
suppose that the initial condition is that it is released from rest from
the triangular position shown in Figure~\ref{Fig02}. Note that the initial
position $f(x,0)$ is not differentiable at $x=L/2$. This is what we call
the problem of the plucked string. The problem is to find $f(x,t)$ for all
$x\in[0,L]$ and all $t\geq 0$. The solution of the problem can be given in
terms of Fourier series for the position, velocity and acceleration of
each point of the string,

\noindent
\begin{eqnarray*}
  f(x,t)
  & = &
  \frac{8h}{\pi^{2}}
  \sum_{j=0}^{\infty}
  \frac{(-1)^{j}}{k^{2}}\,
  \cos\!\left(\pi\frac{k\nu t}{L}\right)
  \sin\!\left(\pi\frac{kx}{L}\right),
  \\
  \frac{\partial f(x,t)}{\partial t}
  & = &
  -\,
  \frac{8h\nu}{\pi L}
  \sum_{j=0}^{\infty}
  \frac{(-1)^{j}}{k}\,
  \sin\!\left(\pi\frac{k\nu t}{L}\right)
  \sin\!\left(\pi\frac{kx}{L}\right),
  \\
  \frac{\partial^{2}f(x,t)}{\partial t^{2}}
  & = &
  -\,
  \frac{8h\nu^{2}}{L^{2}}
  \sum_{j=0}^{\infty}
  (-1)^{j}
  \cos\!\left(\pi\frac{k\nu t}{L}\right)
  \sin\!\left(\pi\frac{kx}{L}\right),
\end{eqnarray*}

\noindent
where $k=2j+1$. Note that the series for the position is absolutely and
uniformly convergent. The series for the velocity can be shown to be
everywhere convergent, but it is not absolutely or uniformly convergent.
The series for the acceleration is simply everywhere divergent. In fact,
in this case it can be shown that it represents two pulses with the form
of Dirac delta ``functions'' going back and forth along the string and
reflecting at its ends. This means that each point of the string is
subjected to repeated impulsive accelerations. These are infinite
accelerations that act for a single instant of time, producing however
finite changes in the velocity. Obviously, we have here a rather singular
situation.

One can change this by applying the first-order linear low-lass filter to
the initial condition, with a range parameter $\varepsilon$, which has the
effect of exchanging the top of the triangle for an inverted arc of
parabola that fits the two remaining segments in such a way that the
resulting function is continuous and differentiable, as shown in
Figure~\ref{Fig02}. Since the coefficients of the equation do not depend
on $x$, we may obtain the filtered solution by simply plugging the filter
factor $[\sin(\pi k\varepsilon/L)/(\pi k\varepsilon/L)]$ into the series,
thus obtaining at once the filtered solution,

\noindent
\begin{eqnarray*}
  f_{\varepsilon}(x,t)
  & = &
  \frac{8hL}{\pi^{3}\varepsilon}
  \sum_{j=0}^{\infty}
  \frac{(-1)^{j}}{k^{3}}\,
  \sin\!\left(\pi\frac{k\varepsilon}{L}\right)
  \cos\!\left(\pi\frac{k\nu t}{L}\right)
  \sin\!\left(\pi\frac{kx}{L}\right),
  \\
  \frac{\partial f_{\varepsilon}(x,t)}{\partial t}
  & = &
  -\,
  \frac{8h\nu}{\pi^{2}\varepsilon}
  \sum_{j=0}^{\infty}
  \frac{(-1)^{j}}{k^{2}}\,
  \sin\!\left(\pi\frac{k\varepsilon}{L}\right)
  \sin\!\left(\pi\frac{k\nu t}{L}\right)
  \sin\!\left(\pi\frac{kx}{L}\right),
  \\
  \frac{\partial^{2}f_{\varepsilon}(x,t)}{\partial t^{2}}
  & = &
  -\,
  \frac{8h\nu^{2}}{L\pi\varepsilon}
  \sum_{j=0}^{\infty}
  \frac{(-1)^{j}}{k}\,
  \sin\!\left(\pi\frac{k\varepsilon}{L}\right)
  \cos\!\left(\pi\frac{k\nu t}{L}\right)
  \sin\!\left(\pi\frac{kx}{L}\right).
\end{eqnarray*}

\noindent
We see that now the series for the position and for the velocity are both
absolutely and uniformly convergent. The series for the acceleration is
now convergent, although it is still not absolutely or uniformly
convergent. It now represents two rectangular pulses of width
$\varepsilon$ propagating back and forth along the string and reflecting
at its ends, with inversion of their sign. This means that each point of
the string is now subjected repeatedly to a large but finite acceleration,
proportional to $1/\varepsilon$, acting for a very short time, of the
order of $\varepsilon/\nu$. One can show that the series for the
acceleration is convergent using trigonometric identities to write it in
the form

\noindent
\begin{eqnarray*}
  \frac{\partial^{2}f_{\varepsilon}(x,t)}{\partial t^{2}}
  & = &
  -\,
  \frac{h\nu^{2}}{L\pi\varepsilon}
  \sum_{j=0}^{\infty}
  \frac{1}{k}\,
  \left[
    \rule{0em}{6ex}
    \hspace{0.7em}
    \sin\!\left(k\pi\frac{L+\varepsilon-\nu t-x}{L}\right)
    +
    \sin\!\left(k\pi\frac{L-\varepsilon+\nu t+x}{L}\right)
  \right.
  +
  \\
  &   &
  \left.
    \rule{0em}{3ex}
    \hspace{6em}
    -
    \sin\!\left(k\pi\frac{L+\varepsilon+\nu t+x}{L}\right)
    -
    \sin\!\left(k\pi\frac{L-\varepsilon-\nu t-x}{L}\right)
  \right.
  +
  \\
  &   &
  \left.
    \rule{0em}{6ex}
    \hspace{6em}
    -
    \sin\!\left(k\pi\frac{L+\varepsilon-\nu t+x}{L}\right)
    -
    \sin\!\left(k\pi\frac{L-\varepsilon+\nu t-x}{L}\right)
  \right.
  +
  \\
  &   &
  \left.
    \rule{0em}{6ex}
    \hspace{6em}
    +
    \sin\!\left(k\pi\frac{L+\varepsilon+\nu t-x}{L}\right)
    +
    \sin\!\left(k\pi\frac{L-\varepsilon-\nu t+x}{L}\right)
    \hspace{1em}
  \right].
\end{eqnarray*}

\noindent
These eight sine series have coefficients that converge monotonically to
zero and therefore are convergent by the Dirichlet test, or alternatively
by the monotonicity criterion discussed in~\cite{FTotCPII}. Therefore, the
series for the acceleration is in fact convergent after the application of
the filter. Note that these eight series represent travelling waves
propagating on an infinite string of which our vibrating string can be
thought of as a given segment.

We can say that the application of the filter in fact {\em improved} the
representation of the physical system in this problem, because it is
unreasonable to imagine that a real physical string could have the initial
format used at first, with the point of non-differentiability. For one
thing, it would be necessary to use some physical object such as a nail or
peg to hold it in its initial position prior to release. The radius of
this object is an excellent candidate for $\varepsilon$. In any case, one
cannot hope to make a perfect angle by bending a material string that has
a finite and non-zero thickness. The radius of the cross-section of the
string would be another excellent candidate for $\varepsilon$. In this way
we see that the application of the filter brought the representation of
the physical system closer to reality.

\subsection{Potential in a Rectangular Box}

\begin{figure}[ht]
  \centering
  \fbox{
    \epsfig{file=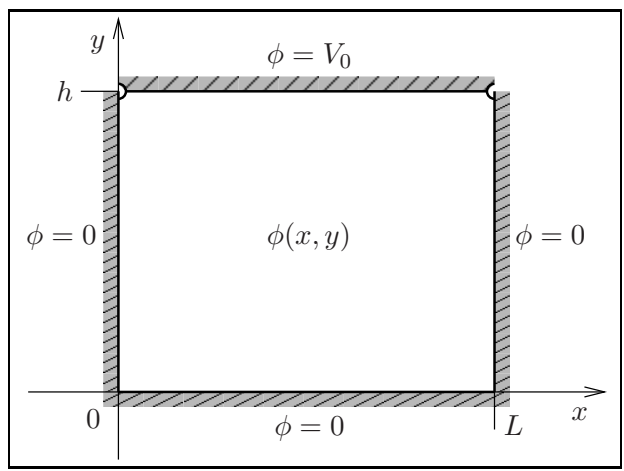,scale=1.0,angle=0}
  }
  \caption{The two-dimensional rectangular box and the boundary values for
    the electric potential.}
  \label{Fig03}
\end{figure}

Consider an empty two-dimensional rectangular box with electrically
conducting walls kept at given values of the electric potential, as shown
on Figure~\ref{Fig03}. The electrostatic potential within the box is given
by Laplace's equation,

\begin{displaymath}
  \frac{\partial^{2}}{\partial x^{2}}\phi(x,y)
  +
  \frac{\partial^{2}}{\partial y^{2}}\phi(x,y)
  =
  0.
\end{displaymath}

\noindent
The boundary values are as shown in the illustration. Note that, as
indicated in the figure, the top surface cannot be considered as being in
electric contact with the other ones, if the potentials are to be kept as
shown. However, in the resolution of the problem this fact is not taken
explicitly into account. The solution of the problem can be given in terms
of mixed Fourier and hyperbolic series for the electric potential and for
the two Cartesian components of the electric field, which is obtained as
minus the gradient of the potential,

\noindent
\begin{eqnarray*}
  \phi(x,y)
  & = &
  \frac{4V_{0}}{\pi}
  \sum_{j=0}^{\infty}
  \frac{1}{k}\,
  \sin\!\left(\pi\frac{kx}{L}\right)\,
  \frac
  {\sinh\!\left(\pi\FFrac{ky}{L}\right)}
  {\sinh\!\left(\pi\FFrac{kh}{L}\right)},
  \\
  E_{x}(x,y)
  & = &
  -\,
  \frac{4V_{0}}{L}
  \Sum_{j=0}^{\infty}
  \cos\!\left(\pi\frac{kx}{L}\right)\,
  \frac
  {\sinh\!\left(\pi\FFrac{ky}{L}\right)}
  {\sinh\!\left(\pi\FFrac{kh}{L}\right)},
  \\
  E_{y}(x,y)
  & = &
  -\,
  \frac{4V_{0}}{L}
  \Sum_{j=0}^{\infty}
  \sin\!\left(\pi\frac{kx}{L}\right)\,
  \frac
  {\cosh\!\left(\pi\FFrac{ky}{L}\right)}
  {\sinh\!\left(\pi\FFrac{kh}{L}\right)},
\end{eqnarray*}

\begin{figure}[ht]
  \centering
  \fbox{
    \epsfig{file=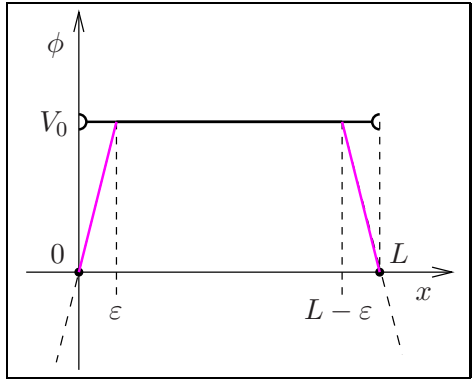,scale=1.0,angle=0}
  }
  \caption{The original boundary condition on the top surface of the box,
    superposed with the filtered boundary condition.}
  \label{Fig04}
\end{figure}

\noindent
where $k=2j+1$. Note that so long as $y<h$ all these series are absolutely
convergent and uniformly convergent along the direction $x$. In fact, in
this case they converge to $C^{\infty}$ functions of $x$. This is so
because the ratios of hyperbolic functions decrease to zero exponentially
fast with $k$ when $y<h$. However, for $y=h$, that is at the top surface,
these factors cease to approach zero as $k\to\infty$, and then the
convergence status is much more precarious. The series for the potential
is still convergent, but not absolute or uniformly so. The series for the
field components diverge everywhere. This is caused by the neglect to take
into account the fact that the top surface must be electrically isolated
from the others, which causes the appearance of infinite electric fields
near the two top corners of the box. This is expressed as a singular
boundary condition at the top surface, with the form of a rectangular
pulse, discontinuous at $x=0$ and $x=L$, as shown in Figure~\ref{Fig04}.

One can change this by applying the first-order linear low-lass filter to
the boundary condition at the top surface, with a small range parameter
$\varepsilon$, which has the effect of exchanging the discontinuities for
two very steep potential ramps, as is also shown in Figure~\ref{Fig04}.
Since the coefficients of the differential equation do not depend on $x$,
we may at once write the solution for the filtered potential and for the
filtered electric field components, by simply plugging the filter factor
$[\sin(\pi k\varepsilon/L)/(\pi k\varepsilon/L)]$ into the series,

\noindent
\begin{eqnarray*}
  \phi_{\varepsilon}(x,y)
  & = &
  \frac{4V_{0}L}{\pi^{2}\varepsilon}
  \sum_{j=0}^{\infty}
  \frac{1}{k^{2}}\,
  \sin\!\left(\pi\frac{k\varepsilon}{L}\right)
  \sin\!\left(\pi\frac{kx}{L}\right)\,
  \frac
  {\sinh\!\left(\pi\FFrac{ky}{L}\right)}
  {\sinh\!\left(\pi\FFrac{kh}{L}\right)}.
  \\
  E_{\varepsilon,x}(x,y)
  & = &
  -\,
  \frac{4V_{0}}{\pi\varepsilon}
  \sum_{j=0}^{\infty}
  \frac{1}{k}\,
  \sin\!\left(\pi\frac{k\varepsilon}{L}\right)
  \cos\!\left(\pi\frac{kx}{L}\right)\,
  \frac
  {\sinh\!\left(\pi\FFrac{ky}{L}\right)}
  {\sinh\!\left(\pi\FFrac{kh}{L}\right)},
  \\
  E_{\varepsilon,y}(x,y)
  & = &
  -\,
  \frac{4V_{0}}{\pi\varepsilon}
  \sum_{j=0}^{\infty}
  \frac{1}{k}\,
  \sin\!\left(\pi\frac{k\varepsilon}{L}\right)
  \sin\!\left(\pi\frac{kx}{L}\right)\,
  \frac
  {\cosh\!\left(\pi\FFrac{ky}{L}\right)}
  {\sinh\!\left(\pi\FFrac{kh}{L}\right)}.
\end{eqnarray*}

\noindent
The series for the potential is now absolutely and uniformly convergent
everywhere within the box, including the top surface. The series for the
field components are still strongly convergent to $C^{\infty}$ functions
away from the top surface, and at that surface they are convergent almost
everywhere, although not absolutely or uniformly so. The solution changes
significantly only in the neighborhood of the two points were the original
singularity on the boundary conditions was located. If we write, as an
example, the field component $E_{\varepsilon,x}(x,y)$ at the top surface,
we get

\begin{displaymath}
  E_{\varepsilon,x}(x,h)
  =
  -\,
  \frac{4V_{0}}{\pi\varepsilon}
  \sum_{j=0}^{\infty}
  \frac{1}{k}\,
  \sin\!\left(\pi\frac{k\varepsilon}{L}\right)
  \cos\!\left(\pi\frac{kx}{L}\right),
\end{displaymath}

\noindent
where $k=2j+1$. One can show that this series is everywhere convergent
using trigonometric identities to write it in the form

\begin{displaymath}
  E_{\varepsilon,x}(x,h)
  =
  \frac{2V_{0}}{\pi\varepsilon}
  \sum_{j=0}^{\infty}
  \frac{1}{k}\,
  \left[
    \sin\!\left(\pi k\frac{x-\varepsilon}{L}\right)
    -
    \sin\!\left(\pi k\frac{x+\varepsilon}{L}\right)
  \right].
\end{displaymath}

\noindent
The two sine series obtained in this way have coefficients that converge
monotonically to zero and therefore are convergent by the Dirichlet test,
or alternatively by the monotonicity criterion discussed
in~\cite{FTotCPII}. Therefore, the series for $E_{\varepsilon,x}(x,y)$ is
in fact everywhere convergent after the application of the filter. The
other field component at the top surface can be analyzed in a similar way.
It is given by

\begin{displaymath}
  E_{\varepsilon,y}(x,h)
  =
  -\,
  \frac{4V_{0}}{\pi\varepsilon}
  \sum_{j=0}^{\infty}
  \frac{1}{k}\,
  \sin\!\left(\pi\frac{k\varepsilon}{L}\right)
  \sin\!\left(\pi\frac{kx}{L}\right)\,
  \frac
  {\cosh\!\left(\pi\FFrac{kh}{L}\right)}
  {\sinh\!\left(\pi\FFrac{kh}{L}\right)}.
\end{displaymath}

\noindent
For large values of $k$ the ratio of hyperbolic functions tends to $1$.
In this case the analysis with the monotonicity criterion shown that the
series is convergent at all points except two, the points $x=\varepsilon$
and $x=L-\varepsilon$. Further application of the first-order filter, or
the application of the second-order filter, can then be used to further
improve the situation.

We can say that the introduction of the filter in fact {\em improved} the
representation of the physical system in this problem, from the physical
standpoint, because the two steep potential ramps can be understood as a
representation of the electric potential within two thin slices of an
insulating material, of thickness $\varepsilon$, inserted between the top
surface and the two lateral ones. In fact, within a good dielectric this
is a very good representation of the electric potential. Once more we see
that the introduction of the filter brought the description of the
physical system closer to reality.

The two remaining isolated points of divergence are in the field component
normal to the surface, exactly at the point where we have the material
interfaces between the conducting material of the upper wall and the
insulating dielectric of the thin slices. This suggests that these
remaining divergences are related to the imperfect representation of these
material interfaces. In fact, the inclusion of the thin insulating slices
introduces into the system two electric capacitors, and the divergences
may be related to the known edge effects that occur at the edges of the
plates of any capacitor, where electric charges tend to accumulate.

In this case the physical scale involved is that of the inter-material
transition at the interface, which may go right down to the molecular
level. Therefore it seems appropriate to use once more the first-order
filter on the boundary condition, this time with a range parameter
$\varepsilon'\ll\varepsilon$. This will have the effect of smoothing out
the sharp transitions between the two materials. Regardless of the value
of the parameter $\varepsilon'$, this will render all the series
absolutely and uniformly convergent everywhere, since they will then all
have at least a factor of $1/k^{2}$ in their coefficients.

The actual values of the field near the material interfaces will of course
depend on $\varepsilon'$. If it turns out to be possible to measure these
field values well enough, it may be possible to determine an optimal value
of $\varepsilon'$ based on experimental data, which will then establish a
rough model for the description of the material interface, from the
macroscopic point of view. We see therefore that this mathematical
technique may be turned into a tool for probing into aspects of the
structure of the physical world.

\subsection{Heat Conduction in a Cylinder}

\begin{figure}[ht]
  \centering
  \fbox{
    \epsfig{file=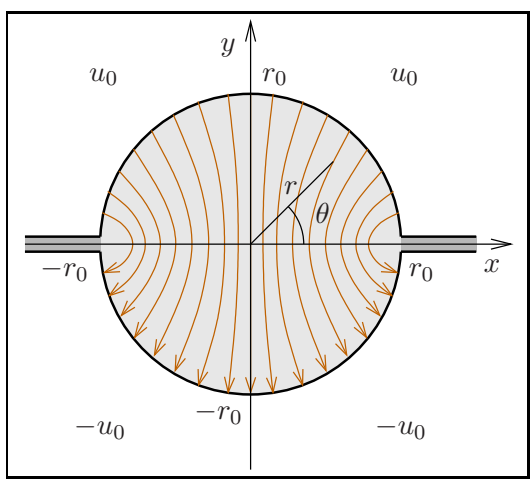,scale=1.0,angle=0}
  }
  \caption{The infinite cylinder and the boundary conditions for the
    temperature.}
  \label{Fig05}
\end{figure}

Consider the two-dimensional cross-section of an infinite solid cylinder
made of a heat-conducting material. Suppose that its two sides are in
contact with heat baths, one above and one below, as shown in
Figure~\ref{Fig05}. The system reaches a state of stationary heat
conduction given by Laplace equation in cylindrical coordinates for the
temperature $u(r,\theta)$,

\begin{displaymath}
  r\,
  \frac{\partial}{\partial r}
  \left[
    r\,
    \frac{\partial}{\partial r}u(r,\theta)
  \right]
  +
  \frac{\partial^{2}}{\partial\theta^{2}}u(r,\theta)
  =
  0.
\end{displaymath}

\noindent
The two boundary values are shown in the illustration. In the standard
(and simpler) formulation of the problem, one considers that the two heat
baths are thermally isolated from each other, but the thermally isolating
material needed to accomplish this is not taken into account explicitly.
The solution of the problem can be given in terms of mixed power and
Fourier series for the static temperature and for the radial and angular
components of the heat flux density, which is related to the gradient of
the temperature,

\noindent
\begin{eqnarray*}
  u(r,\theta)
  & = &
  \frac{4u_{0}}{\pi}
  \sum_{j=0}^{\infty}
  \frac{1}{k}
  \left(\frac{r}{r_{0}}\right)^{k}
  \sin(k\theta),
  \\
  \jmath_{r}(r,\theta)
  & = &
  -\,
  \frac{4c\mu\kappa u_{0}}{\pi r_{0}}
  \sum_{j=0}^{\infty}
  \left(\frac{r}{r_{0}}\right)^{k-1}
  \sin(k\theta),
  \\
  \jmath_{\theta}(r,\theta)
  & = &
  -\,
  \frac{4c\mu\kappa u_{0}}{\pi r_{0}}
  \sum_{j=0}^{\infty}
  \left(\frac{r}{r_{0}}\right)^{k-1}
  \cos(k\theta),
\end{eqnarray*}

\begin{figure}[ht]
  \centering
  \fbox{
    \epsfig{file=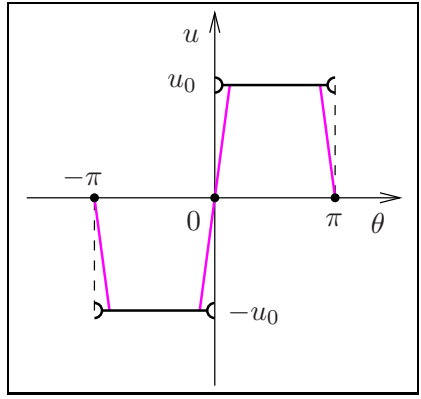,scale=1.0,angle=0}
  }
  \caption{The original boundary conditions superposed with the filtered
    boundary conditions.}
  \label{Fig06}
\end{figure}

\noindent
where $k=2j+1$ and the constants appearing in $\jmath_{r}(r,\theta)$ and
$\jmath_{\theta}(r,\theta)$ characterize the material. Note that so long
as $r<r_{0}$ all the series are strongly convergent to $C^{\infty}$
functions, due to the exponential decay with $k$ of the factors involving
the ratio $(r/r_{0})$. However, at the surface of the cylinder, for
$r=r_{0}$, the series for the temperature is not absolutely or uniformly
convergent, but only point-wise convergent. Besides, at this surface the
series for the components of the heat flux density are simply everywhere
divergent. This is caused by the neglect to take into account the
necessity to have a layer of isolating material of finite thickness
between the two heat baths, which causes the presence of infinite heat
fluxes at the points where these two heat baths are infinitely close to
each other, and connected to each other through the material of the
cylinder.

One can change this by applying the first-order linear low-pass filter to
the boundary condition at the surface of the cylinder, with a small
angular range parameter $\epsilon$. This exchanges the two discontinuities
of the boundary temperature for two thin layers with steep variation of
the temperature, as shown in Figure~\ref{Fig06}. Since the coefficients of
the differential equation do not depend on $\theta$, we may write at once
the filtered solution, by simply plugging the filter factor
$[\sin(k\epsilon)/(k\epsilon)]$ into the series,

\noindent
\begin{eqnarray*}
  u_{\epsilon}(r,\theta)
  & = &
  \frac{4u_{0}}{\pi\epsilon}
  \sum_{j=0}^{\infty}
  \frac{\sin(k\epsilon)}{k^{2}}
  \left(\frac{r}{r_{0}}\right)^{k}
  \sin(k\theta),
  \\
  \jmath_{\epsilon,r}(r,\theta)
  & = &
  -\,
  \frac{4c\mu\kappa u_{0}}{\pi r_{0}\epsilon}
  \sum_{j=0}^{\infty}
  \frac{\sin(k\epsilon)}{k}
  \left(\frac{r}{r_{0}}\right)^{k-1}
  \sin(k\theta),
  \\
  \jmath_{\epsilon,\theta}(r,\theta)
  & = &
  -\,
  \frac{4c\mu\kappa u_{0}}{\pi r_{0}\epsilon}
  \sum_{j=0}^{\infty}
  \frac{\sin(k\epsilon)}{k}
  \left(\frac{r}{r_{0}}\right)^{k-1}
  \cos(k\theta),
\end{eqnarray*}

\noindent
where $k=2j+1$. The series for the temperature is now absolutely and
uniformly convergent everywhere, and the series for the components of the
heat flux density are point-wise convergent almost everywhere at the
surface of the cylinder. If we write for example the components
$\jmath_{\epsilon,\theta}(r,\theta)$ at the surface, we get

\begin{displaymath}
  \jmath_{\epsilon,\theta}(r_{0},\theta)
  =
  -\,
  \frac{4c\mu\kappa u_{0}}{\pi r_{0}\epsilon}
  \sum_{j=0}^{\infty}
  \frac{1}{k}\,
  \sin(k\epsilon)
  \cos(k\theta).
\end{displaymath}

\noindent
One can show that this series is convergent everywhere using trigonometric
identities to write it in the form

\begin{displaymath}
  \jmath_{\epsilon,\theta}(r_{0},\theta)
  =
  \frac{2c\mu\kappa u_{0}}{\pi r_{0}\epsilon}
  \sum_{j=0}^{\infty}
  \frac{1}{k}\,
  \left\{
    \rule{0em}{2.5ex}
    \sin[k(\theta-\epsilon)]
    -
    \sin[k(\theta+\epsilon)]
  \right\}.
\end{displaymath}

\noindent
The two sine series obtained in this way have coefficients that converge
monotonically to zero and therefore are convergent by the Dirichlet test,
or alternatively by the monotonicity criterion discussed
in~\cite{FTotCPII}. The other field component at the surface can be
analyzed in a similar way. It is given by

\begin{displaymath}
  \jmath_{\epsilon,r}(r_{0},\theta)
  =
  -\,
  \frac{4c\mu\kappa u_{0}}{\pi r_{0}\epsilon}
  \sum_{j=0}^{\infty}
  \frac{1}{k}\,
  \sin(k\epsilon)
  \sin(k\theta).
\end{displaymath}

\noindent
In this case the analysis with the monotonicity criterion shown that the
series is convergent at all points except four, the points
$\theta=\pm\epsilon$, $\theta=\pi-\epsilon$ and $\theta=-\pi+\epsilon$.
Further application of the first-order filter, or the application of the
second-order filter, can then be used to further improve the situation.

We can say that the introduction of the filter in fact {\em improved} the
representation of the physical system in this problem, from the physical
standpoint, because the two steep variations of the temperature can be
understood as representations of the temperature within two thin slices of
a thermally insulating material, of angular thickness $2\epsilon$,
inserted between the two heath baths. Once again we see that the
introduction of the filter brought the description of the physical system
closer to reality.

Just an in the previous example, the remaining points of divergence are in
the component of the heat flux density normal to the surface, exactly at
the point where we have the material interfaces between the heat baths and
the thermally insulating material of the thin slices. This once more
suggests that these remaining divergences are related to the imperfect
representation of these material interfaces. Therefore it seems
appropriate to use once more the first-order filter on the boundary
condition, with a range parameter $\epsilon'\ll\epsilon$ which could go
all the way down to the molecular scale.

Although a direct and detailed physical interpretation seems not to be
immediately apparent in this case, essentially the same comments made in
the last example about the role of a second application of the filter are
also true in this case. Clearly it will have the effect of smoothing out
the sharp transitions between the two materials. Regardless of the value
of the parameter $\epsilon'$, it will certainly render all the series
absolutely and uniformly convergent everywhere, since they will then all
have at least a factor of $1/k^{2}$ in their coefficients.

Note that since the derivation of the heat equation from fundamental
physical principles has a statistical character, involving averages over
large numbers of molecules of the material involved, it is not
unreasonable that one may meet with difficulties in its description of
nature when one goes down to the molecular scale, as we have done here.
We may interpret these isolated singularities as consequences of the use
of a physical theory at the very edge of its recognized domain of
validity.

\end{document}